\documentclass[12pt,preprint]{aastex}

\shorttitle{The BSS population in NGC 5824}
\shortauthors{N. Sanna et al.}

\begin{document} 

\title{The WFPC2 UV Survey: the BSS population in NGC 5824
  \thanks{Based on observations with the NASA/ESA {\it HST}
    (Prop. 11975), obtained at the Space Telescope Science Institute,
    which is operated by AURA, Inc., under NASA contract NAS5-26555.}}

\author{
N.~Sanna\altaffilmark{1,2},
E.~Dalessandro\altaffilmark{2},  
F.R.~Ferraro\altaffilmark{2}, 
B.~Lanzoni\altaffilmark{2}, 
P.~Miocchi\altaffilmark{2}, 
R.~W.~O'Connell\altaffilmark{3}
}

\affil{\altaffilmark{1} INAF-Osservatorio Astrofisico di Arcetri, largo Fermi 5, 50127 Firenze, Italy\\
$E-mail: sanna@arcetri.astro.it$}
\affil{\altaffilmark{2} Dipartimento di Fisica e Astronomia, Universit\`a degli Studi di Bologna, via Ranzani 1, I--40127 Bologna, Italy}
\affil{\altaffilmark{3} Department of Astronomy, University of Virginia, P. O. Box 400325, Charlottesville, VA 22904, USA}

\begin{abstract}
We have used a combination of high-resolution \textit{Hubble Space
  Telescope} WFPC2 and wide-field ground-based observations, in
ultraviolet and optical bands, to study the blue straggler star
population of the massive outer-halo globular cluster NGC 5824, over
its entire radial extent.
We have computed the
center of the cluster and constructed the radial density
profile, from detailed star counts. The profile is well reproduced by
a Wilson model with a small core ($r_c \simeq 4.4\arcsec$) and a
concentration parameter $c \simeq 2.74$. We also present the first
age determination for this cluster. From the comparison with
isochrones, we have found $t=13\pm0.5$ Gyr. We discuss this result in
the context of the observed age-metallicity relation of Galactic
globular clusters.
A total of 60 bright blue stragglers has
been identified. Their radial distribution is found to be bimodal,
with a central peak, a well defined minimum at $r \sim 20\arcsec$, and
an upturn at large radii. In the framework of the dynamical clock
defined by Ferraro et al. (2012), this feature suggests that NGC 5824
is a cluster of intermediate dynamical age.
\end{abstract} 


\section{INTRODUCTION}

The high stellar densities typical of globular clusters (GCs) make stellar interactions 
very likely events. For this reason it is expected that in GCs stellar evolution 
is strongly affected by the
environment. Indeed GCs are efficient furnaces of exotic populations, like X-ray binaries, 
millisecond pulsars and blue straggler stars (BSSs; Paresce et al. 1992; Bailyn 1995; Bellazzini et al. 1995;
Ferraro et al. 1995, 2001; Ransom et al. 2005; Pooley \& Hut 2006).
BSSs constitute the largest population among these $''$exotica$"$,  thus representing a crucial probe 
of the GC internal dynamics. In the optical color-magnitude diagram (CMD) of
stellar systems BSSs appear bluer and brighter than the main
sequence (MS) stars, thus mimicking a population of objects
younger and more massive than the normal MS turnoff stars.
Indeed, direct measurements (Shara et al. 1997, Gilliland et al. 1998,
De Marco et al. 2004) have shown that BSSs are two to three times more
massive than the average stars in GCs ($\langle m\rangle\sim 0.5
M_\odot$).

Two main scenarios have been proposed to explain the formation of
BSSs: mass transfer in binary systems (McCrea 1964; Zinn $\&$ Searle
1976) and stellar collisions (Hills $\&$ Day 1976). The two formation
channels are believed to co-exist within the same cluster (see for
example the case of M30 Ferraro et al.  2009; Li et al. 2013; Dalessandro et al. 2013) 
with efficiencies that
may vary as a function of the environment (Bailyn 1992, Ferraro et al
1995). Collisional BSSs more likely form in the core or in very dense
GCs because the densities are higher and so are the probabilities of
direct collisions (e. g. Mapelli et al. 2004).  
Mass transfer is expected to be the dominant
formation channel in the cluster's outskirts or in loose environments
(but see Knigge et al. 2009). 
Chemical signatures of the different BSS
formation mechanisms have been found in 47 Tucanae (Ferraro et
al. 2006a) and in M30 (Lovisi et al. 2013) where a fraction of stars
shows anomalies in carbon and oxygen abundances, which are expected to
be signatures of the mass-transfer process (Sarna $\&$ de Greve 1996).

Because of their mass and the mechanisms involved in their formation,
BSSs are powerful tracers of the evolution of internal dynamics of
clusters. In particular, as recently shown by Ferraro et al. (2012,
F12), their radial distribution provides us with a measure of the
dynamical friction efficiency. In fact, based on the observed shape of
several BSS distributions, F12 grouped GCs in three main
families. Family I is composed of clusters where the radial BSS
distribution is indistinguishable from that of the reference stars,
suggesting that these clusters have not undergone mass-segregation yet
($\omega$ Centauri, Ferraro et al. 2006b; NGC 2419, Dalessandro et
al. 2008a; Palomar 14, Beccari et al. 2011). Most of the clusters show
a bimodal BSS radial distribution: a peak in the center, a clear dip
at intermediate radii, and an upturn in the external regions. These
clusters are members of Family II.  Family III is composed by clusters
with a monotonically decreasing BSS distribution (M79, Lanzoni et
al. 2007a; M75, Contreras Ramos et al. 2012; M80 and M30, F12). In
these clusters even the most distant BSSs already drifted toward the
cluster's center. Different families correspond to different dynamical
age of the clusters. As proposed by F12, Family I systems are
dynamically young, Family II clusters have intermediate dynamical ages
and the Family III ones are the oldest.

Being more massive than the average cluster stars, BSSs are typically
found in the innermost regions of GCs, where stellar crowding and the
dominant luminosity contribution of red giant branch (RGB) stars make
the construction of complete samples of BSSs quite difficult in
optical bands. Conversely, it is fairy easy in the UV bands (Paresce
et al. 1991).  In particular, high-resolution UV observations with the
\textit{Hubble Space Telescope} (HST) open the possibility to survey
UV bright populations, including horizontal branch (HB; Ferraro et
al. 1998; D'Cruz et al. 2000; Dalessandro et al. 2011; see also
O'Connell et al. 1997 and Schiavon et al. 2012), BSSs (Ferraro et
al. 1993; Ferraro et al. 1997) and post-asymptotic giant branch (see
for example Brown et al. 2008) stars, even in the innermost regions.

In the framework of an extensive UV survey of more than 30 Galactic
GCs conducted with the Wide Field Planetary Camera 2 (WFPC2) on board
HST (Prop 11975, PI Ferraro), here we present a detailed
multiwavelength photometric analysis of NGC 5824.  This poorly studied
cluster is located at $\sim26$ kpc from the Galactic center (Harris
1996, 2010 version) and, after NGC 2419, it is the most luminous
outer-halo GC ($M_V=-8.85$, Harris 1996).  Newberg et al. (2009)
suggested that NGC 5824 might be associated to the Cetus Polar Stream
and it could have once been a dwarf galaxy core (Georgiev et
al. 2009).  Using the $Ca_{II}$ triplet measured for 17 RGB stars,
Saviane et al. (2012) found a possible intrinsic iron spread of
$\sigma([Fe/H]=0.12)$ dex in this cluster. Grillmair et al. (1995)
found that the surface density profile follows a power law over almost
the entire extent of the cluster.  Lutzgendorf et al. (2013) have
recently suggested the presence of a $2000M_{\odot}$ black hole in
this clusters.

The paper is organized as follows. Section 2 describes the data sets
and the photometric and astrometric analysis. In Section 3 we
determine the center of gravity and the radial density profile of the
system. In Section 4 we present the first age determination for this
cluster. In Sections 5 and 6 we discuss the BSS properties and present
our conclusions. The summary of the paper is presented in Section 7.

\section{OBSERVATIONS AND DATA ANALYSIS}
As in our previous studies (e.g. Sanna et al. 2012 and references
therein), we have used a combination of high resolution and wide-field
data to resolve the stars in the central regions and to cover the
entire radial extension of the cluster at the same time. We used the
best quality HST and ground-based data available for this cluster.

\begin{figure}
\includegraphics[scale=0.83]{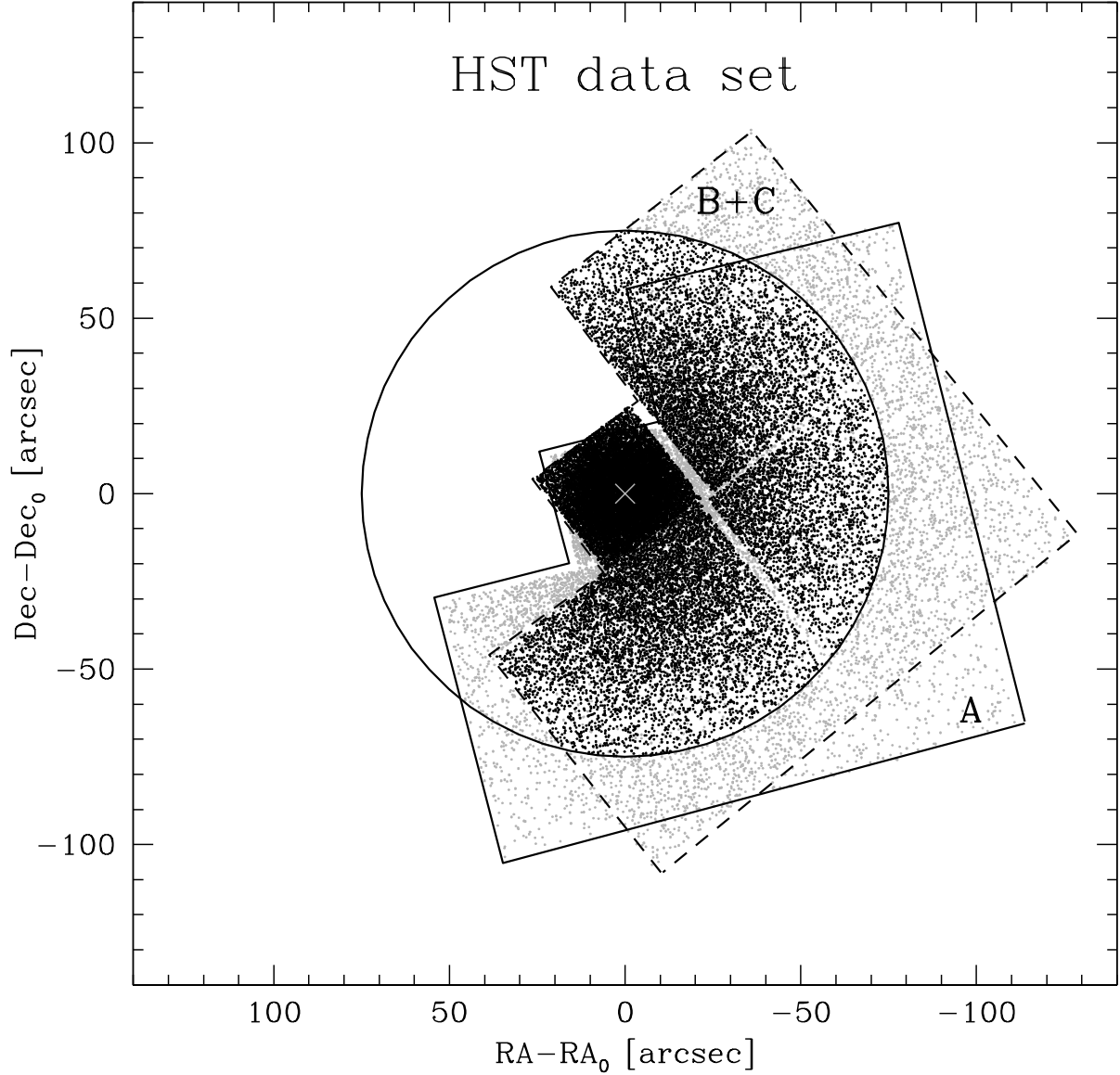}
\caption{Map of the HST data set. The solid line marks the FOV of
  pointing A, the dashed line marks the FOV of pointings B and C, the
  grey cross indicates the cluster center $C_{\rm grav}$.  The
  \textit{inner sample} (see Sect. \ref{sec_center}) consists of all
  the stars (black dots) included in the FOV of pointings B+C and
  within a distance $r=75\arcsec$ from $C_{\rm grav}$ (solid circle).}
\label{map_hst}
\end{figure}

The \textit{HST data set} consists of a series of images collected
with three different pointings of the WFPC2 in several bands, ranging
from the UV to the optical.  Pointing A includes 11 optical and UV
images (Prop 11975, PI Ferraro) obtained through the filters F255W, 
F336W and F555W with total exposure times $t_{\rm exp}=7200$\,s, 
$t_{\rm exp}= 2700$\,s and $t_{\rm
  exp}=200$\,s, respectively.  Pointing B consists of 12 optical
images obtained through the filters F336W, F439W and F555W (Prop 5902,
PI Fahlman) with total exposure times $t_{\rm exp}= 1200$\,s, $t_{\rm
  exp}= 3200$\,s and $t_{\rm exp}=300$\,s, respectively.  Pointing C
includes three images obtained through the filter F439W (Prop 8095, PI
Ibata) with total exposure time $t_{\rm exp}= 1200$\,s. Pointings B
and C exactly overlap each other. The fields of view (FOVs) of the
three pointings are shown in Figure \ref{map_hst}. The center of the
cluster is located in the Planetary Camera (pixel scale $\sim
0.05\arcsec$ pixel$^{-1}$) in all pointings (see Figure
\ref{map_hst}).

\begin{figure}
\includegraphics[scale=0.83]{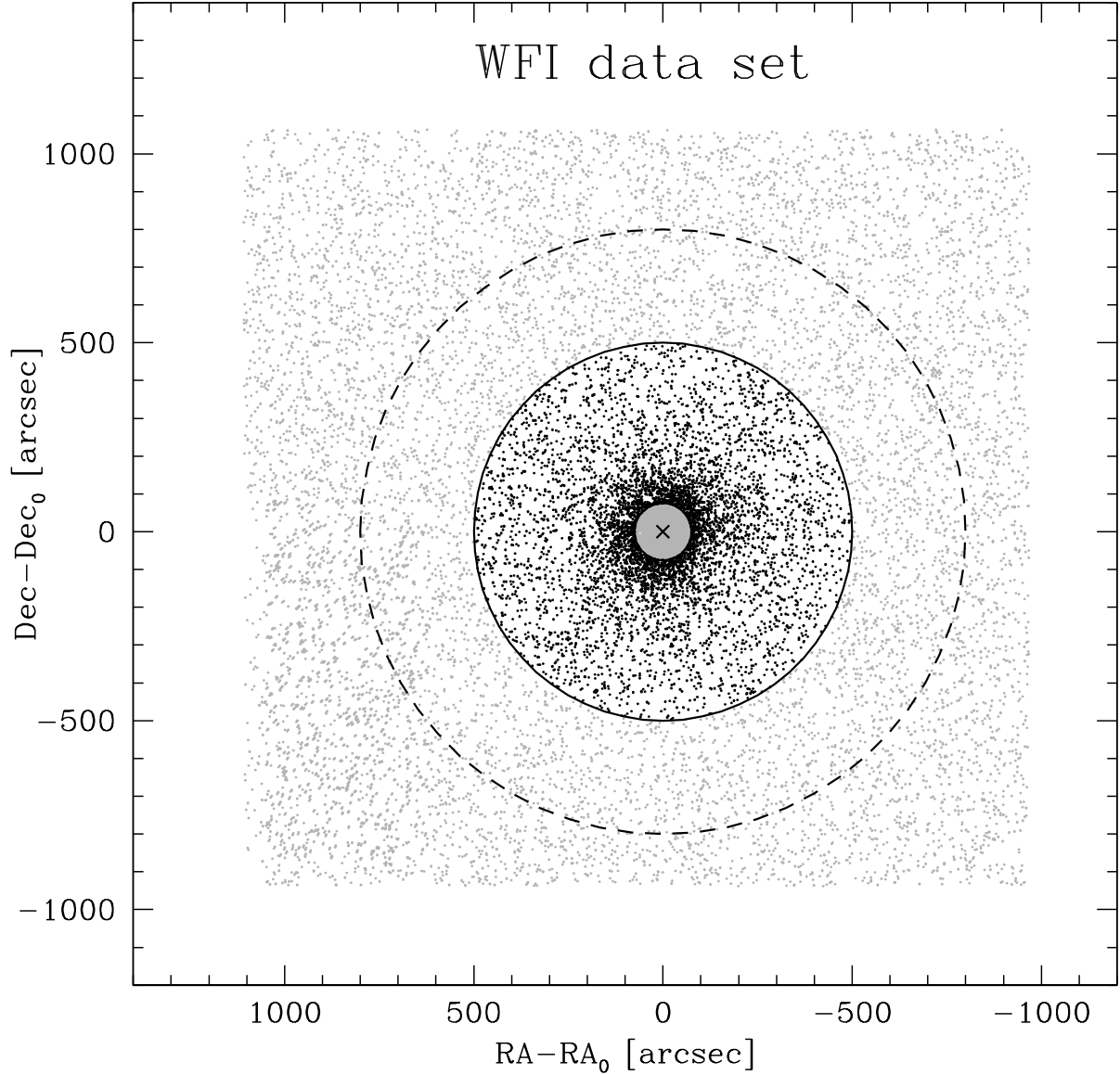}
\caption{Map of the WFI data set. The \textit{outer sample} (see
  Sect. \ref{sec_center}) consists of all the stars observed at
  $r>75\arcsec$ (most internal circle). The radial distribution of the
  various stellar populations (BSSs, RGB and HB stars) has been
  studied only for $r<500\arcsec$ (black dots within the large solid
  circle), where the cluster density becomes comparable to that of
  field stars. To estimate the Galactic field contamination, we have
  used the stars observed at $r>800\arcsec$ (dashed circle). Only
  stars with $m_{555}<22$ are shown for the sake of clarity.}
\label{map_wfi}
\end{figure}

The \textit{WFI data set} is composed of data obtained with the Wide
Filed Imager (WFI) at the 2.2 m ESO/MPI telescope. Two images per
filter through the $B$ and $V$ bands with total exposure times $t_{\rm
  exp}=3600$\,s and $t_{\rm exp}=500$\,s, respectively, were retrieved
from the ESO/STECF Science Archive. WFI consists of 8 CCDs with a
pixel scale of $\sim 0.238\arcsec$ pixel$^{-1}$. The $33' \times 34'$
field of view allowed a complete sampling of the cluster. The cluster
center is located in chip $\#$7 (see Figure \ref{map_wfi}).

The data reduction has been performed by using DAOPHOT IV package
(Stetson 1987, 1994) for both the HST and the WFI data sets.
The analysis of the HST images has been performed on the individual
frames.  As a first step we modeled the point spread function (PSF) by
using $\sim15$ bright and almost isolated stars in each
frame. Typically the same stars were chosen in each filter.  With the
obtained PSF models we performed a first PSF-fit on each single image
by using ALLSTAR.  As second step, we built a median image for each
filter using IRAF tools\footnote{IRAF is distributed by he National
  Optical Astronomy Observatory, which is operated by the Association
  of Universities for Research in Astronomy, Inc., under cooperative
  agreement with the National Science Foundation.}. Then for each
median image we built a list of stars detected above a given
background threshold. The lists thus obtained have been combined for
each chip.  For pointings A and B we built a master-list frame
including only stars detected in at least two median images. With this
criterion, and for pointing A in particular, we secured a master-list
complete for both hot (UV bright) and cold (UV faint) stars.  The
stars in the master-list have then been force-fitted to each single
frame by using the ALLFRAME package (Stetson 1994). This procedure has
been performed separately for pointing A, while it has been applied
simultaneously to the images in pointings B and C.  Unfortunately,
because of the poor quality of the optical images of pointing A, we
used only those of pointings B and C.  Therefore the final
high-resolution HST catalogue includes all the stars located in the
pointings B and C.  UV magnitudes have been assigned to the stars in
common with pointing A.

All the raw WFI images were corrected for bias and flat-fields using
standard IRAF tools. All the single images were reduced using the
DAOPHOT IV package. To model the PSF, typically 100 bright and almost
isolated stars in each frame have been selected. The PSF-fitting on
each single image was performed by using ALLSTAR. Using
{\tt{DAOMATCH/DAOMASTER}} packages the photometry obtained for each
single chip has been reported to a common reference frame (Stetson
2000) that we have chosen to be a $50' \times 60'$image obtained
combining different ground-based data. Once all the chips were on the
same reference frame, we built a master list composed by all the stars
detected in at least two frames. These stars have then been forced to
each single frame by using the ALLFRAME package.  This choice allowed
us to get full advantage of the dithering strategy adopted for these
observations and to completely fill the gaps between different chips.

All the catalogues were put on the absolute astrometric system using
more than 10000 stars in common with the \textit{Guide Stars
  Catalogue} (GSC2.3). As a first step we obtained the astrometric
solution for the entire WFI catalogue by using the procedure described
in Ferraro et al. (2001, 2003) and the cross-correlation tool
CataXcorr (Montegriffo, private communication).  The HST (x, y)
coordinates were first transformed to the HST World Coordinate System
coordinates and then they were reported to the absolute astrometric
system by using the stars in common with WFI.  At the end of the
procedure the estimated error in the absolute positions, both in right
ascension ($\alpha$) and declination ($\delta$), is of about $0.2''$.

All the WFPC2 magnitudes ($m_{255}$, $m_{336}$, $m_{439}$
and $m_{555}$) were calibrated to the VEGAMAG system by using the
prescription by Holtzman et al. (1995) and the zero points from the
WFPC2 data
handbook.\footnote{http://documents.stsci.edu/hst/wfpc2/documents/handbooks/dhb}
We used the equations by Dolphin (2009) to correct for charge transfer
efficiency. We converted the $B$ and $V$ magnitudes of the wide-field
catalogue to the $m_{439}$ and $m_{555}$ VEGAMAG system,
respectively, by means of the following color equations:
$m_{439}=B+0.500\times(B-V)+8.800$, $m_{555}=V-0.037\times(B-V)+4.980$,
obtained by using the stars in common between the two catalogues.

\begin{figure}
\includegraphics[scale=0.83]{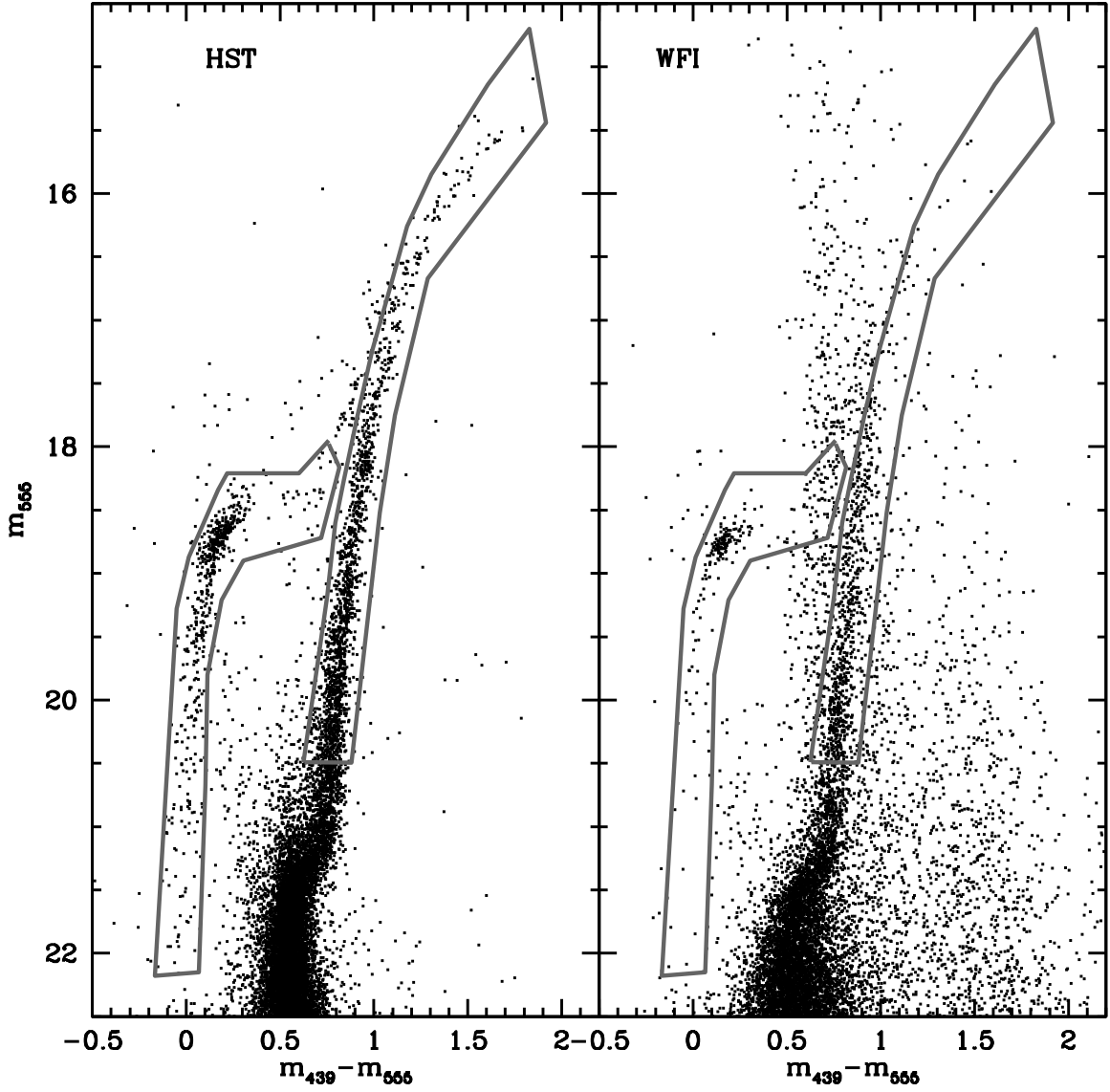}
\caption{Optical CMDs of NGC 5824 for the \textit{inner sample}
  (left panel) and the \textit{outer sample} (right panel).  Only
  the stars included in the grey boxes and with $m_{555}<20$ have
  been used to determine the density profile.}
\label{opt_cmd}
\end{figure}

In order to exclude extra-galactic sources from our analysis, we
matched our catalogues with the NASA EXTRAGALACTIC
DATABASE.\footnote{http://ned.ipac.caltech.edu/} We found 35 objects
in common (typically galaxies and quasars) which were excluded from
the following analysis.  The optical CMDs for the HST and WFI data
sets are shown in Figure  \ref{opt_cmd}.

\section{CENTER OF GRAVITY AND RADIAL DENSITY PROFILE}
\label{sec_center}
We determined the center of gracity ($C_{\rm grav}$) of NGC$ $5824
from resolved bright stars taking advantage of the high-resolution of 
the WFPC2 data. We followed the iterative procedure described by
Montegriffo et al. (1995), as already done in previous studies (see
for example Sanna et al.  2012).
We used the center reported by Harris
(1996) as first guess of our iterative procedure. We averaged the
positions $\alpha$ and $\delta$ of the stars contained within circles
of three different radii ($8''$, $10''$ and $12''$), until convergence
was reached. We were of course limited in this analysis by the size
and the shape of the Planetary Camera FOV. In order to avoid any
possible spurious effect due to incompleteness of the catalogue, we
considered three samples with different limiting magnitudes ($m_{555}=
19.5, 19.8, 20.0$). The obtained values agreed within $\sim0.2''$ and
their average was therefore assumed as $C_{\rm grav}$: $\alpha
(J2000.0)=15^{\rm h} 3^{\rm m} 58.64^{\rm s}$, $\delta
(J2000.0)=-33^{\rm \circ} 4' 5.905''$. This new determination is
located $\sim 0.32''$ South-West ($\Delta\alpha\simeq-0.09''$,
$\Delta\delta\simeq-0.31''$) from the Harris center.

Starting from $C_{\rm grav}$, we divided the entire data set in two
sub-samples. From the HST data set we selected only stars with a
distance $r<75''$ from $C_{\rm grav}$ (see Figure \ref{map_hst}): this
is named the \textit{inner sample}, composed of 22242 stars. Even
if the region at $r<75''$ is not entirely sampled by the WFPC2 FOV, we
conservatively preferred not to complement it with ground-based data.
Hence, from the WFI data set we considered all stars at $r>75''$ (see
Figure \ref{map_wfi}), thus defining the \textit{outer sample}
consisting of 48828 objects.

\begin{figure}
\includegraphics[scale=0.83]{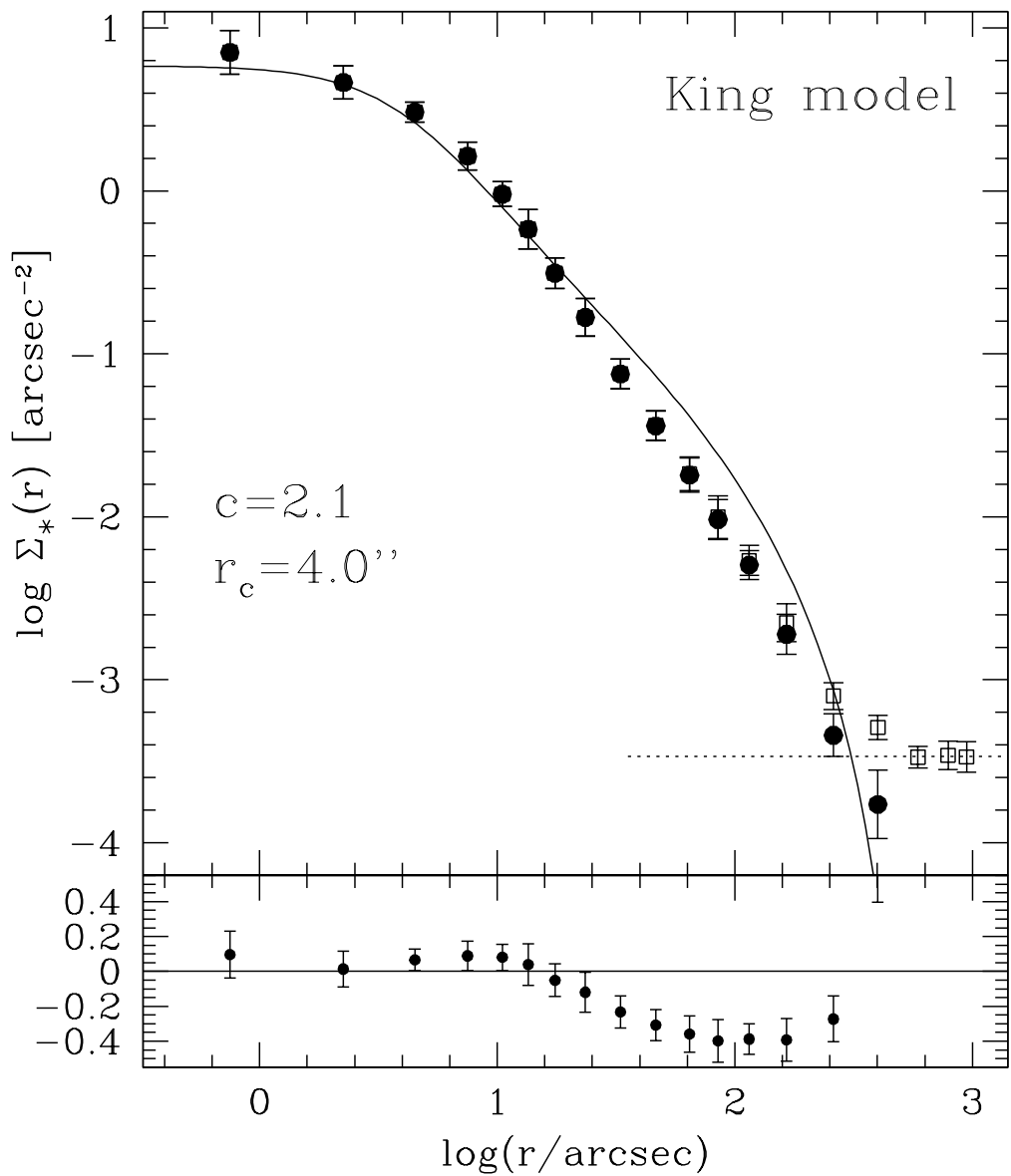}
\caption{Observed surface density profile (empty squares) in units of
  number of stars per square arcseconds. The dotted line indicates the
  adopted level of the Galactic background, computed as the average of
  the three outermost points. Solid circles show the
  background-subtracted density profile. The best-fit King model is
  plotted as a solid line and its residuals with respect to the
  observations are shown in the lower panel. The labels quote the
  values of the model concentration and core radius.}
\label{king}
\end{figure}

\begin{figure}
\includegraphics[scale=0.83]{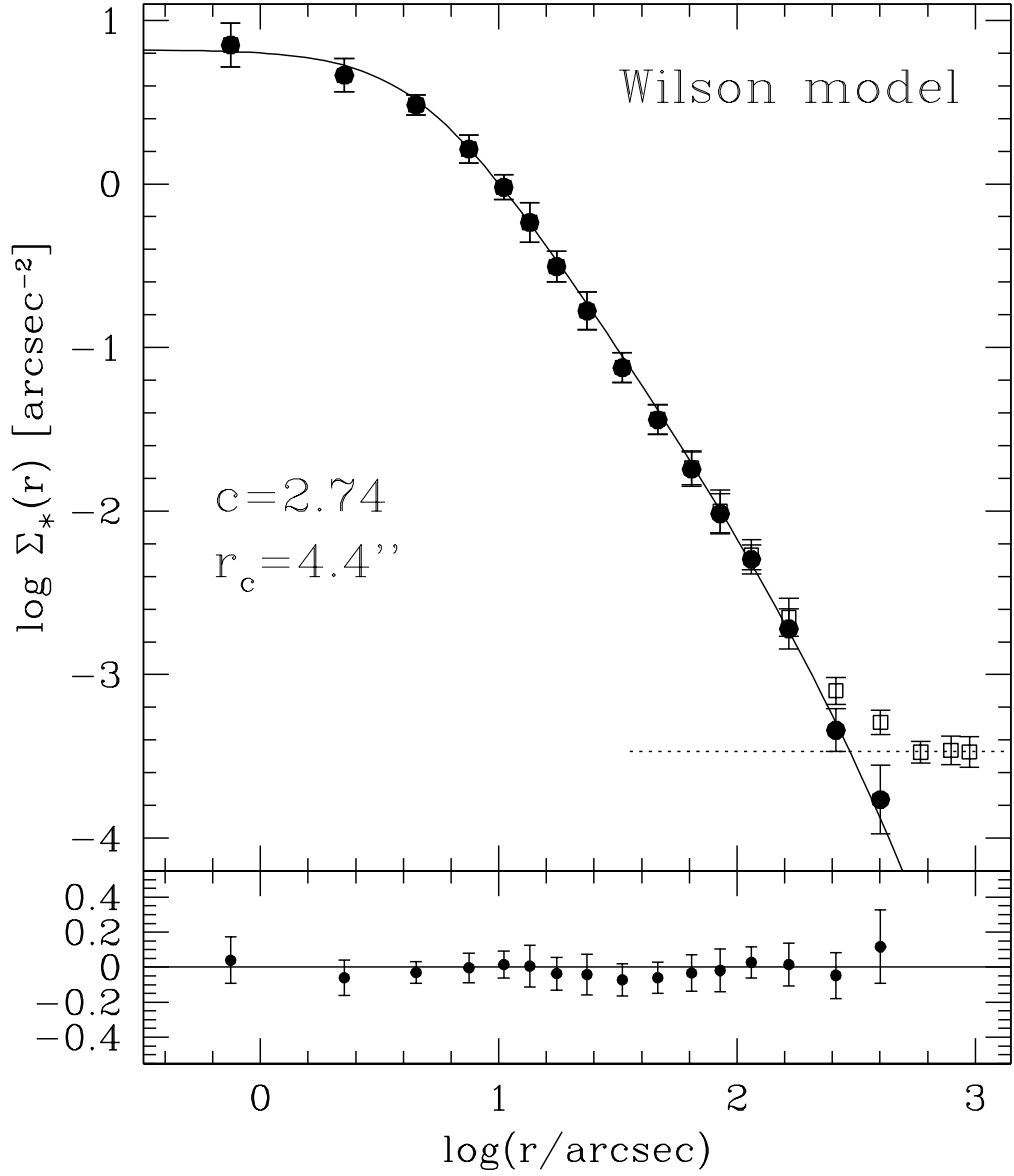}
\caption{As in Figure 4, but for the best-fit Wilson model.}
\label{wilson}
\end{figure}

We determined the projected density profile of NGC 5824 by measuring
the star counts in concentric annuli covering the entire cluster
extension, from $C_{\rm grav}$ to $r \sim 1050''$. In order to limit
the strong contamination from field stars in the most external
regions, only fiducial RGB and HB stars (see Figure \ref{opt_cmd})
have been taken into account. Starting from this first selection and
because of the high crowding affecting the very central regions
($r\sim1.5''$), we further limited our selections to stars with
$m_{555}<20.0$. We divided the entire FOV in 19 annuli centered on
$C_{\rm grav}$ and each annulus was divided into two, three or four
sub-sectors. In each sub-sector the ratio between the number of stars
and the sampled area has been computed. The stellar density of each
annulus is then obtained as the average of the corresponding
sub-sector densities and the associated errors is computed from the
squared root of their variance. Incomplete area coverage affecting
some annuli has been properly taken into account in this procedure.
The resulting radial density profile is shown in Figures \ref{king}
and \ref{wilson}.  The three outermost annuli ($r>500''$) have been
used to estimate the contribution of background stars.

We have tried to reproduce the observed density profile by using both
King (King 1966) and Wilson (Wilson 1975) models.  These are widely
used to describe stellar systems like GCs that are thought to have
reached a state of (quasi-)equilibrium. The projected density profile
in both cases is characterized by a constant value in the innermost
portion and a decreasing behavior outward, with the Wilson model
showing a more extended outer region (for more details see, e.g.,
Miocchi et al. 2013).  We find that the observed density profile
cannot be reproduced by a standard mono-mass King model.  Similar
conclusions have been reached by Grillmair et al. (1995) using
photographic photometry and star counts.  Figure \ref{king} shows the
best fit obtained by using a King model: as can be seen, in the
external regions ($r>30''$) the observed star density profile shows a
radial decrease significantly steeper than predicted by the model.
The profile is instead nicely reproduced by a single mass Wilson model
with a concentration $c\simeq2.74$ and core radius (i.e., the radius
at which the central surface density equals its central value)
$r_c\simeq4.4''$ (see Figure \ref{wilson}). The nominal limiting
radius (at which the model density drops to zero) is located at $r\sim2700''$.  
These values are only marginally
consistent with those ($c=2.87$, $r_c=3.88''$) quoted by McLaughlin \&
van der Marel (2005) for the Wilson model fit to fit the surface
brightness (instead of the surface density) profile.

\section{METALLICITY SPREAD AND AGE DETERMINATION}
As quoted in Section 1, Saviane et al. (2012) recently suggested that
NGC 5824 shows a possible metallicity spread of the order of 0.1 dex.
The position and the morphology of the RGB is a sensible function of
the metallicity (see, e.g., Valenti et al. 2004). Thus, a metallicity
spread is expected to produce a measurable dispersion in the color
distribution of the RGB in the ($m_{555}, m_{439}-m_{555}$) CMD and we
can use our data to put constrains on the possible metallicity spread.

\begin{figure}
\includegraphics[scale=0.83]{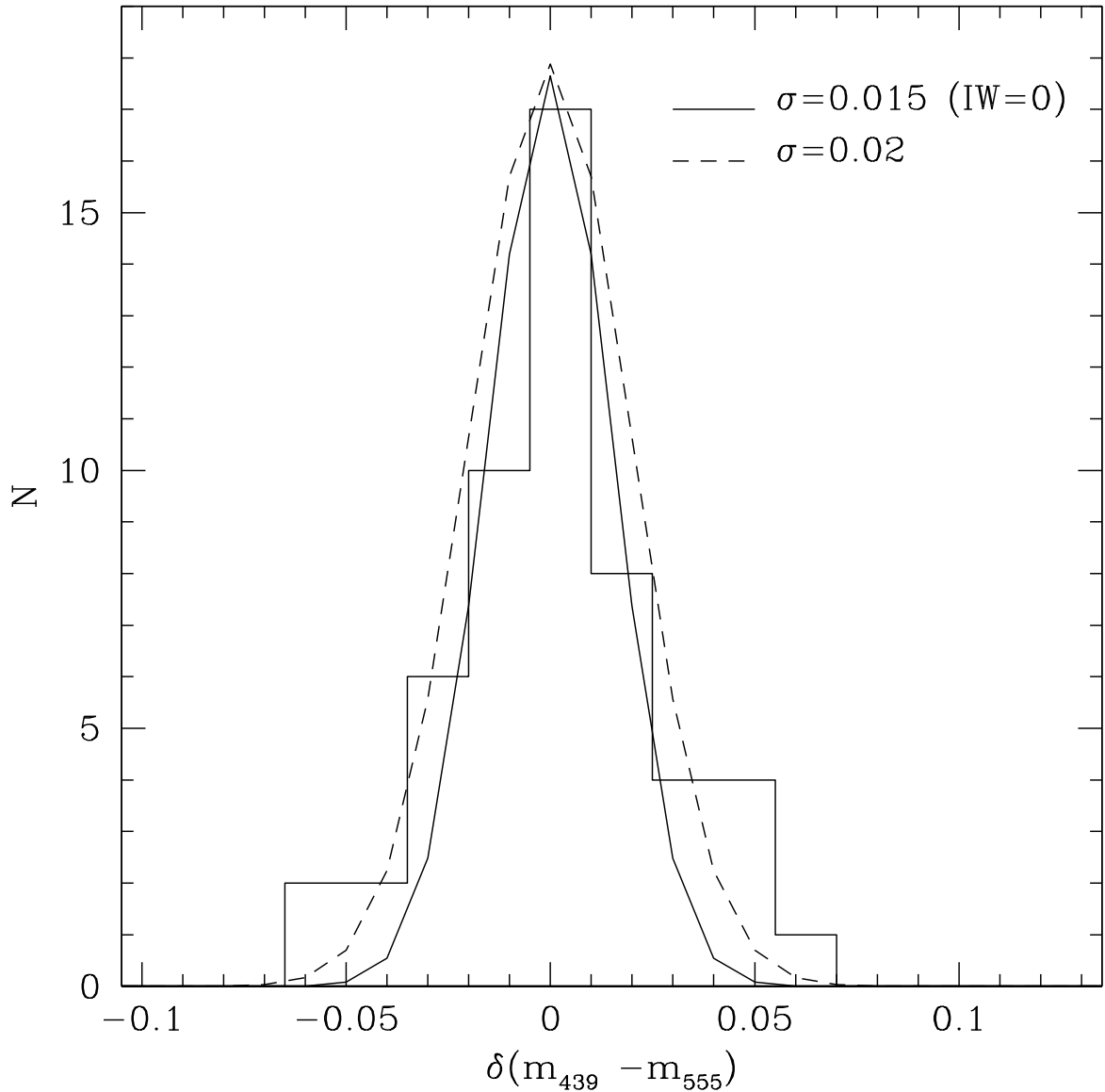}
\caption{Observed color width of the RGB for the stars detected in the
  Wide Field 3 chip of the WFPC2, along the RGB at $18\le m_{555}\le
  20.5$ and at $r<75\arcsec$.  The observed distribution of the
  $(m_{439}-m_{555})$ color residuals with respect to the RGB mean
  ridge line is shown as an histogram. The solid line represents the
  distribution of the internal photometric errors (a Gaussian with
  $\sigma=0.015$), the dashed line corresponds to the same
  distribution convolved with a Gaussian with dispersion
  $\sigma=0.02$, simulating a metallicity spread of $\delta$[Fe/H]=0.1
  dex.}
\label{met}
\end{figure}

For a quantitative estimate, we followed the procedure described in
previous papers (e.g., Ferraro et al. 1991, 1992) to determine the
intrinsic width (IW) of the RGB. In order to minimize at most any
possible bias that can artificially broaden the color distribution of
the RGB, we selected the most vertical portion of the RGB
($18<m_{555}<20.5$) from all stars at $r<75\arcsec$ observed with the
Wide Field 3 chip of the WFPC2.  We then computed the distribution of
the residuals in the $(m_{439}-m_{555})$ color with respect to the
adopted RGB mean ridge line (see the histogram in Figure
\ref{met}). This has been compared with the distribution of the
intrinsic photometric errors (corresponding to IW=0, solid line), and
with such a distribution convolved with a metallicity spread of 0.1
dex (dashed line; note that according with theoretical isochrones,
$\delta(m_{439}-m_{555})/\delta$[Fe/H]=0.0125).  Based on a
$\chi^2$-test, the solid line turns out to better reproduce the
observed distribution, meaning that the color distribution of the
selected RGB stars is consistent with no metallicity spread (IW=0) or
with a spread smaller than 0.1 dex.

\begin{figure}
\includegraphics[scale=0.83]{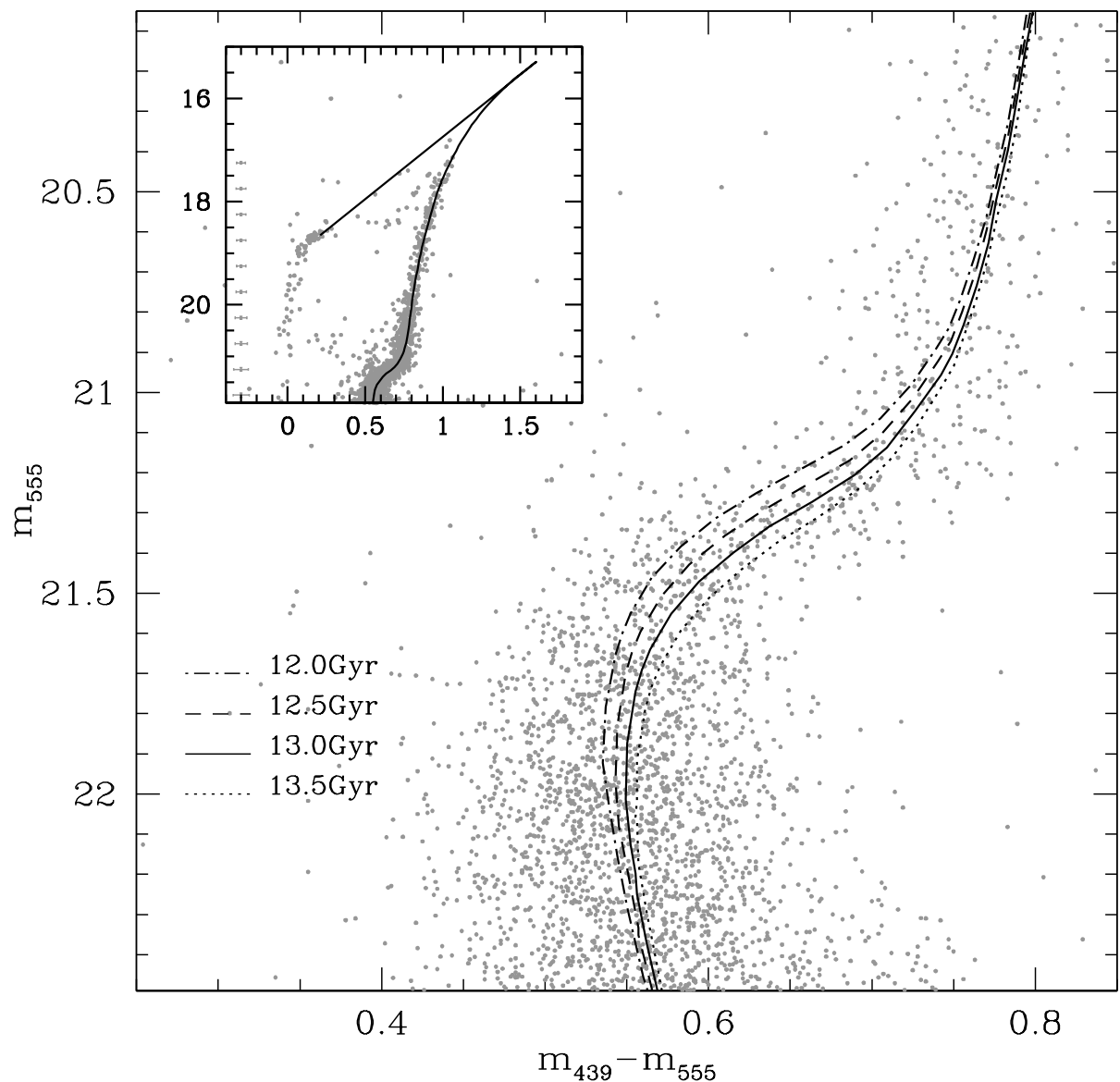}
\caption{Optical CMD of the \emph{inner sample} zoomed in the
  MS-turnoff region, for $30\arcsec\leq r \leq75\arcsec$. Superimposed
  are isochrones from the Girardi's database computed at the cluster
  metallicity ($Z=0.000187$, [Fe/H]$=-1.91$) and at different ages
  (see labels).  The isochrone that best fits the observed CMD is the
  13 Gyr one (solid line). The inset shows an enlarged view of the
  CMD.}
\label{age}
\end{figure}

\begin{figure}
\includegraphics[scale=0.83]{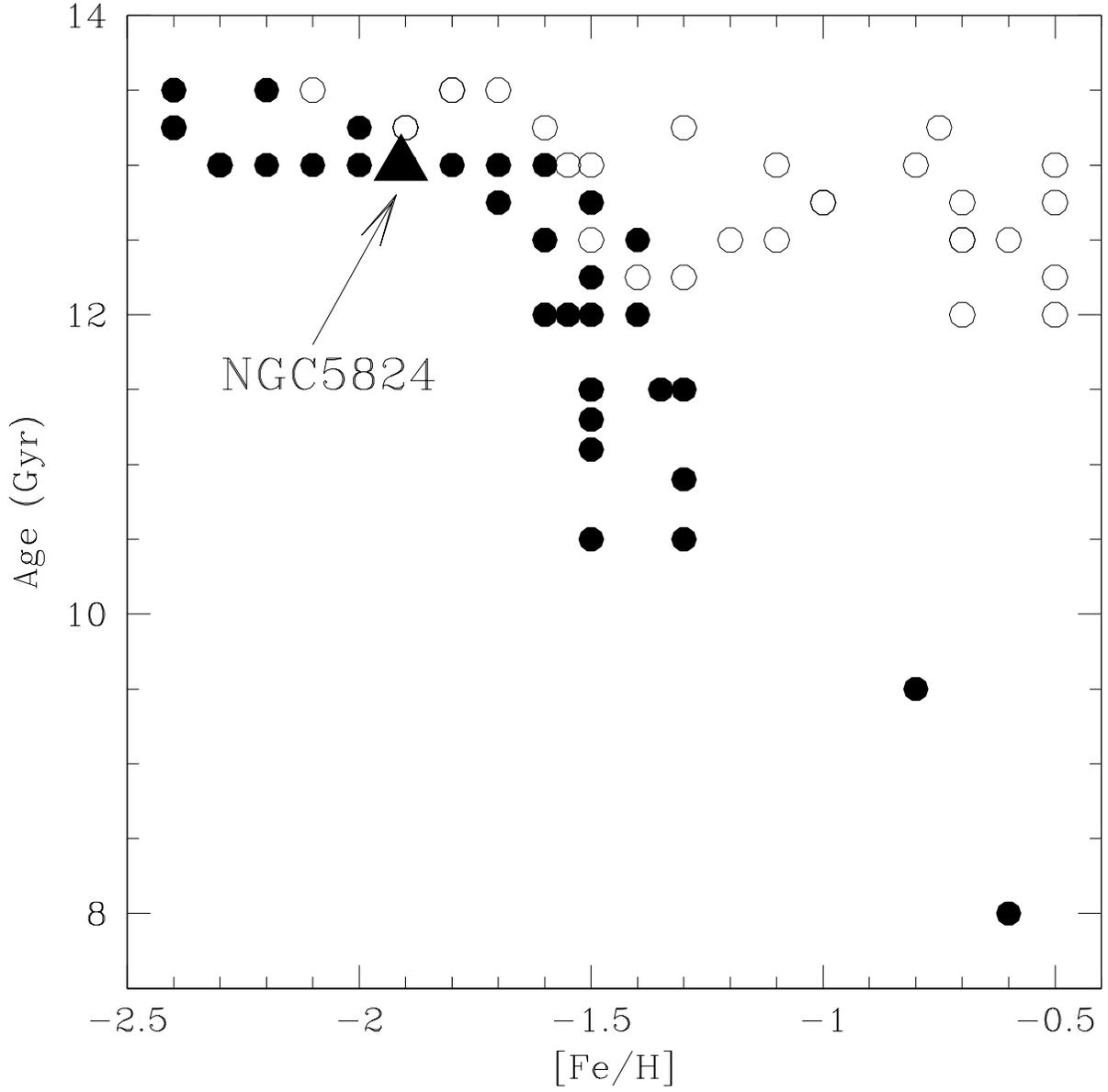}
\caption{Age-metallicity relation for Galactic GCs at $R_{\rm GC}<8$
  kpc (open circles) and $R_{\rm GC}>8$ kpc (filled circles) from
  Dotter et al. (2011). The position of NGC 5824 is marked by the
  large filled triangle.}
\label{age_met}
\end{figure}

The collected data set also offers the opportunity to determine for
the first time the age of NGC 5824.  This is of great importance in
the context of the formation scenarios of our Galaxy. In fact, studies
focused on the age-metallicity relation of Galactic GCs (e.g., Dotter
et al. 2010, 2011) show that two different formation histories can be
distinguished: a rapid chemical enrichment for the GCs in the inner
regions of the Galaxy (at a galactocentric distance $R_{\rm GC}<8$
kpc), and a prolonged formation for those in the outskirts.  In these
studies, however, there is a lack of clusters between $\sim 20$ and
$\sim 40$ kpc from the Galactic center, both because they are few and
because the available photometric data are typically not accurate
enough.  In this context, the case of NGC 5824, which lies at $R_{\rm
  GC}\sim 26$ kpc (Harris 1996) is quite interesting.  In order to
determine the age of this cluster, we have compared the optical HST
CMD with isochrones from the Girardi's database (Bressan et
al. 2012). For a more accurate determination, we have excluded the
first $30''$, where the photometric quality is lower. We have
adopted a distance modulus $(m-M)_0=17.53$ and a reddening
$E(B-V)=0.14$ (Ferraro et al. 1999).  The assumed metallicity is
[Fe/H]$=-1.91$ (Harris 1996).  We have superimposed to the CMD four
isochrones with different ages, ranging between 12.0 Gyr and 13.5 Gyr,
stepped by 0.5 Gyr (see Figure \ref{age}).  From this comparison we
have determined $t=13\pm0.5$ Gyr (solid line).  As shown in Figure
\ref{age_met}, the derived age puts NGC 5824 in nice agreement with
the age-metallicity relation discussed by Dotter et al (2010, 2011).
NGC 5824 is in the metallicity regime where the two samples of
Galactic GCs (at $R_{\rm GC}<8$ kpc and $R_{\rm GC}>8$ kpc) share the
same portion of the age-metallicity relation plane.

\section{THE BSS POPULATION}
As already done in previous studies (e.g. Lanzoni et al. 2007b;
Dalessandro et al. 2008b; Sanna et al. 2012, and references therein),
in order to perform a meaningful study of BSSs in terms of both
specific frequency and radial distribution we need to select a
population of "normal" cluster stars, like HB or RGB stars, which are
expected to be distributed as the cluster light and, for this reason,
represent a reference population.

\subsection{The BSS selection}
As quoted in the Introduction, UV-CMDs are the best planes to study
the hottest stellar populations in GCs. For this reason we selected
the BSSs in the ($m_{255},m_{255}-m_{336}$) CMD.  In order to minimize
possible contamination from the MS turnoff and sub-giant branch, we
limited the sample to $m_{255}<21.6$ and we also applied a selection
in color: $0<(m_{255}-m_{336})<1.1$. The adopted selection box is
shown in Figure \ref{uv_sel}. Within these limits, we identified 37
BSSs from the UV.

Unfortunately, the UV data do not cover the entire extension of the
cluster. Hence, as done in previous papers (see Lanzoni et al. 2007b) 
we "translated" the UV selection box in the
optical plane in the following way:  we identified the position of the UV-selected BSSs in
the ($m_{555},m_{439}-m_{555}$) HST CMD, then we defined the boundaries of the optical selection box 
as to include the bulk of the UV-selected BSSs. The resulting optical
selection box is shown in Figure \ref{opt_sel}. From the portion of
the \textit{inner sample} not covered by pointing A we thus
selected 3 stars, for a total of 40 BSSs in the \textit{inner
  sample}.

The same box has been used to selected BSSs in the \textit{outer
  sample}.  In this case, we limited the analysis to $r<500\arcsec$,
the distance at which the cluster density becomes smaller than the
background density (see Figure \ref{wilson}). Even if this value is
several times smaller than the estimated limiting radius (see Section 3), we
conservatively preferred to adopt this limit to minimize the impact of
contamination for Galactic field stars.  In the \textit{outer sample}
we then selected 23 BSSs.

\subsection{The reference populations}
\begin{figure}
\includegraphics[scale=0.83]{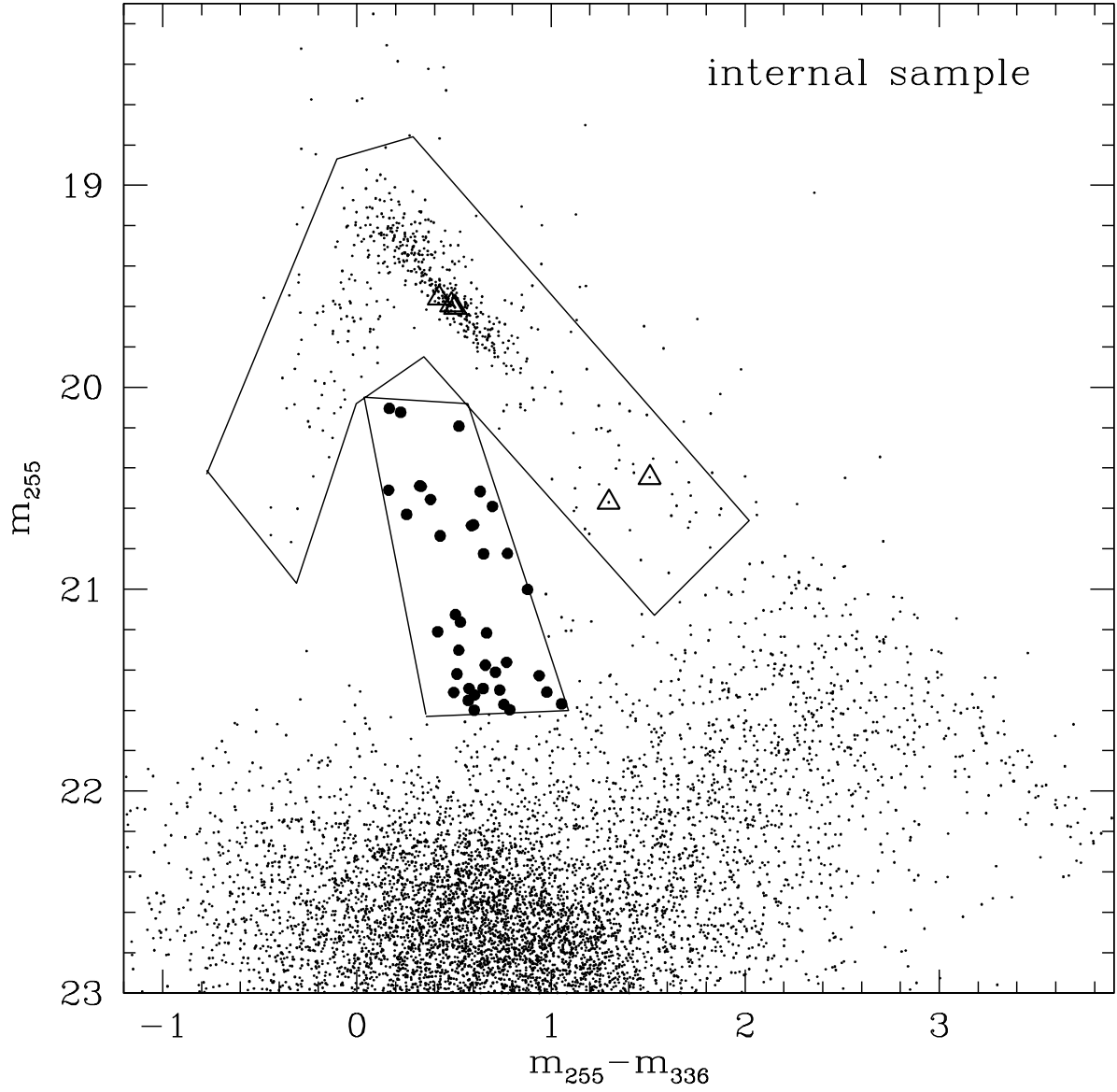}
\caption{UV CMD of the \emph{inner sample}. The adopted BSS and HB
  selection boxes are shown. Solid circles highlight the selected
  BSSs, while empty triangles mark the known RR Lyrae stars.}
\label{uv_sel}
\end{figure}

\begin{figure}
\includegraphics[scale=0.83]{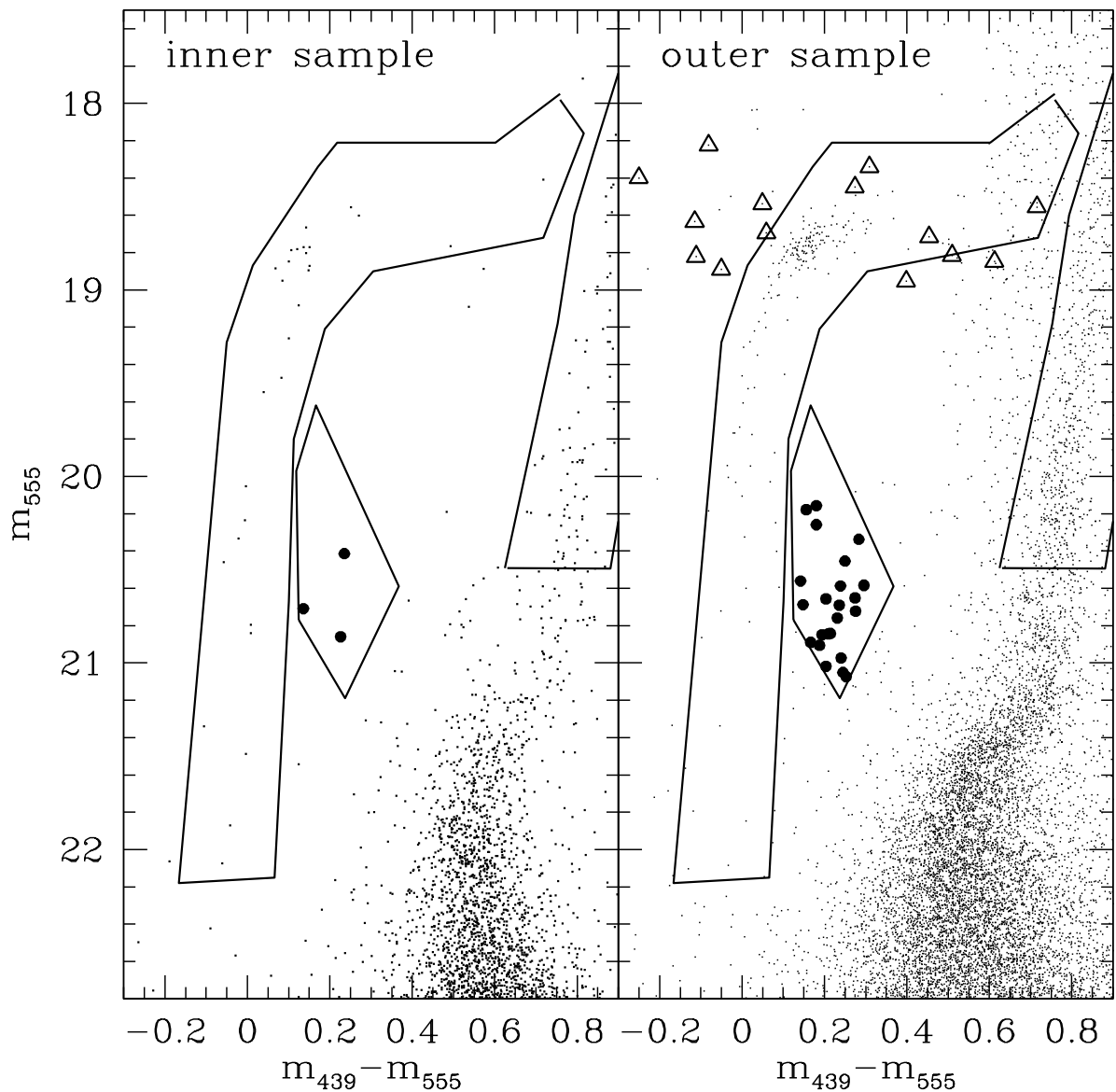}
\caption{Optical CMDs of NGC 5824 zoomed in the BSS region. The left
  panel shows the CMD of the \textit{inner sample} not covered by
  the UV data; the CMD of the \textit{outer sample} for $r\le
  500\arcsec$ is shown in the right-hand panel.  The adopted BSS and
  HB selection boxes are shown with solid lines. Solid circles mark
  the BSSs, while empty triangles mark the known RR Lyrae stars.}
\label{opt_sel}
\end{figure}

As representative of the normal cluster stars we considered both the
HB and the RGB populations.  Since NGC 5824 has an extended HB, the UV
diagram is the best plane to select these stars.  Following the same
procedure adopted for BSSs, we defined the HB selection box in the UV
plane (Figure \ref{uv_sel}) and we then converted this selection in
the optical plane.  NGC 5824 hosts 26 confirmed RR Lyrae stars (Samus
et al. 2009), 19 of which are located in the FOV covered by our
data (five in the \textit{inner sample} and 14 in the \textit{outer sample}). 
These have been included in the HB selection (see triangles in
Figures \ref{uv_sel} and \ref{opt_sel}).  Considering the entire
cluster, from $C_{\rm grav}$ to $r=500''$, we identified 819 HB stars,
557 in the \textit{inner sample} and 262 in the complementary
\textit{outer sample}.

To select the RGB stars we used the optical CMDs, where these objects
are bright and the branch well defined.  Following the RGB mean ridge
line, we limited our selection to $m_{555}<20.5$ (see the selection
boxes in Figure \ref{opt_cmd}). We thus identified 2286 RGB stars,
1478 in the \textit{inner sample} and 808 in the \textit{outer
  sample}.

\subsection{The BSS radial distribution}
\begin{figure}
\includegraphics[scale=0.83]{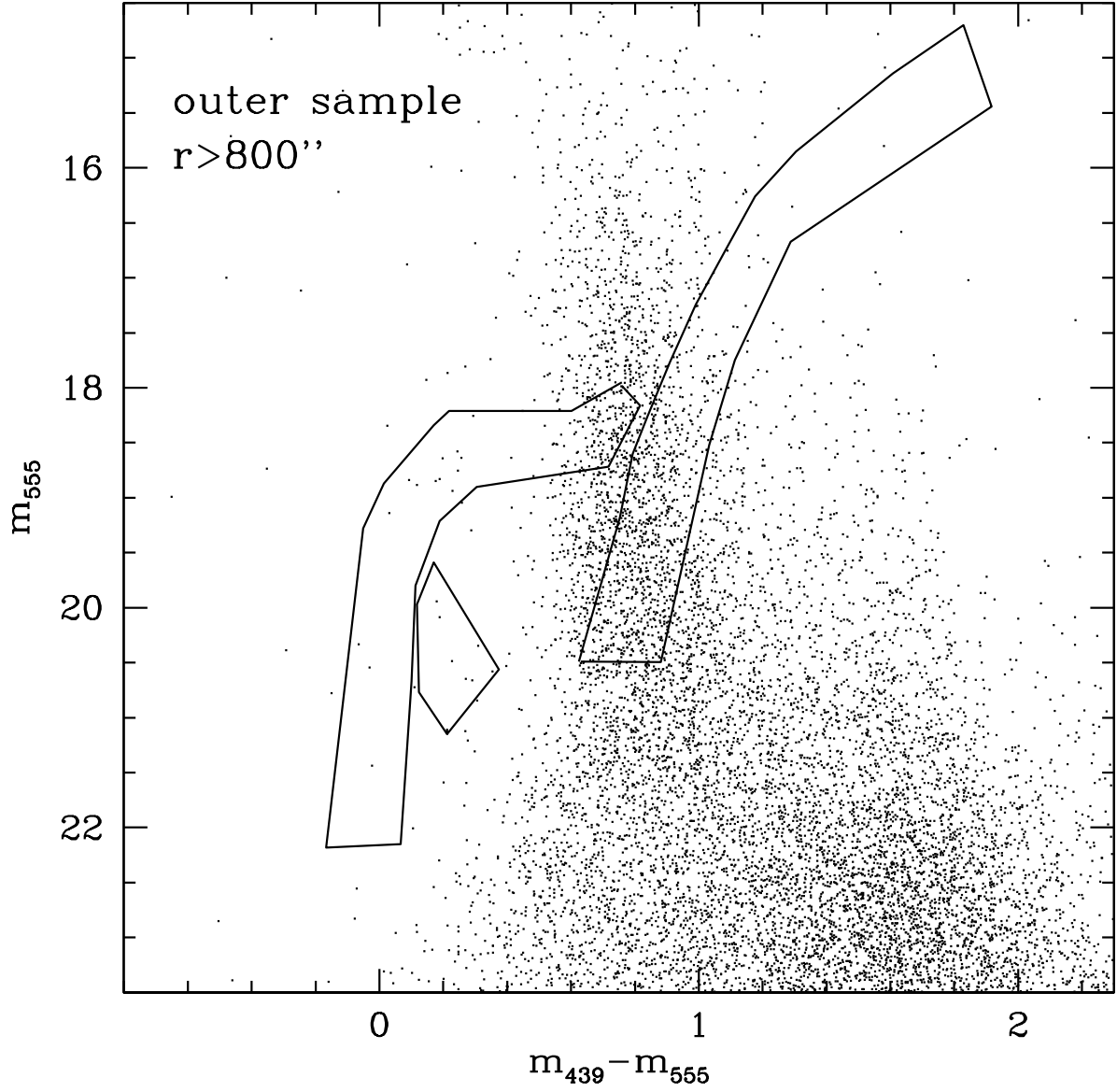}
\caption{CMD of the \emph{outer sample} at $r>800\arcsec$, used to
  estimate the contamination of Galactic field stars to the BSS, HB
  and RGB population selections (the corresponding selection boxes are
  marked with solid lines).}
\label{field}
\end{figure}

As evident from the CMD shown in the right panel of Figure
\ref{opt_cmd}, the WFI data set is strongly contaminated by field
stars.  For this reason we carefully estimated the expected number of
field stars in each selection box.  In order to statistically quantify
the Galactic field contamination we used the CMD obtained for
$r>800''$, where field stars define two vertical sequences roughly
located at $0.5<(m_{439}-m_{555})<1$ and $1.4<(m_{439}-m_{555})<1.8$
(see Figure \ref{field}).  By counting the number of stars located
within the boxes used for the population selections, we derived the
following values for the field star densities: $\rho_{\rm
  {BSS}}\sim0.0131$ stars arcmin$^{-2}$, $\rho_{\rm {HB}}\sim0.2732$
stars arcmin$^{-2}$, $\rho_{\rm {RGB}}\sim1.3023$ stars arcmin$^{-2}$.
Star counts after decontamination are: $N_{\rm BSS}=60$, $N_{\rm
  HB}=759$, $N_{\rm RGB}=2004$.

\begin{figure}
\includegraphics[scale=0.83]{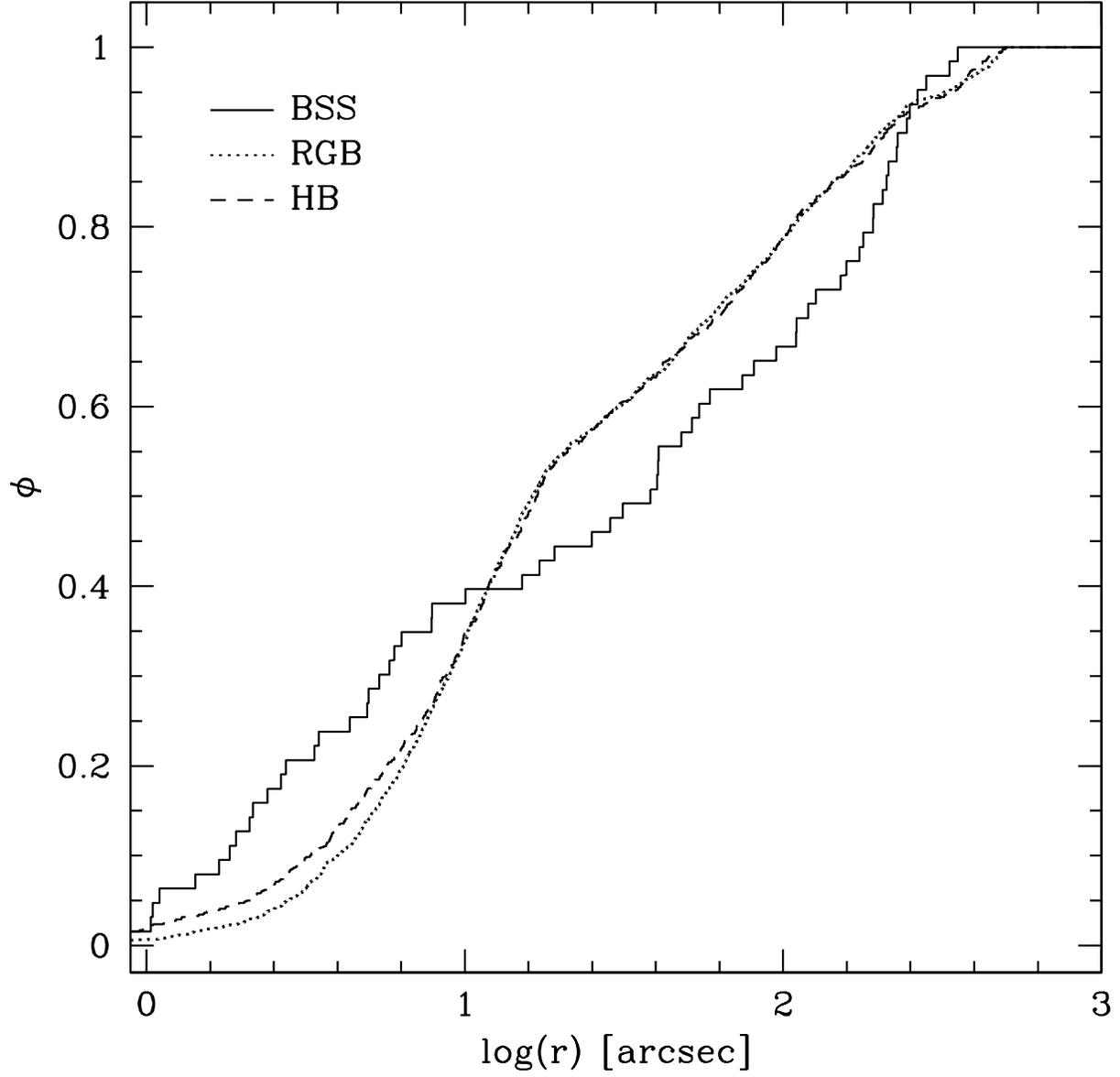}
\caption{Cumulative radial distribution of the statistically
  decontaminated populations of BSSs (solid line), HB stars (dashed
  line) and RGB stars (dotted line) as a function of the projected
  distance from $C_{\rm grav}$.}
\label{ks}
\end{figure}

Following Dalessandro et al. (2013), we divided the FOV into five
concentric annuli centered on $C_{\rm grav}$ and, for each of them, we
randomly subtracted a number of stars constrained by the computed
field star densities.  Figure \ref{ks} shows the statistically
decontaminated cumulative radial distribution for the three selected
populations. As evident, BSSs are more centrally concentrated than RGB
and HB stars. The Kolmogorov-Smirnov test gives a probability of
$\sim0.08$ and $\sim0.18$ that the BSS distribution is extracted from
the same parent distribution as the RGB and HB stars, respectively.

\begin{figure}
\includegraphics[scale=0.83]{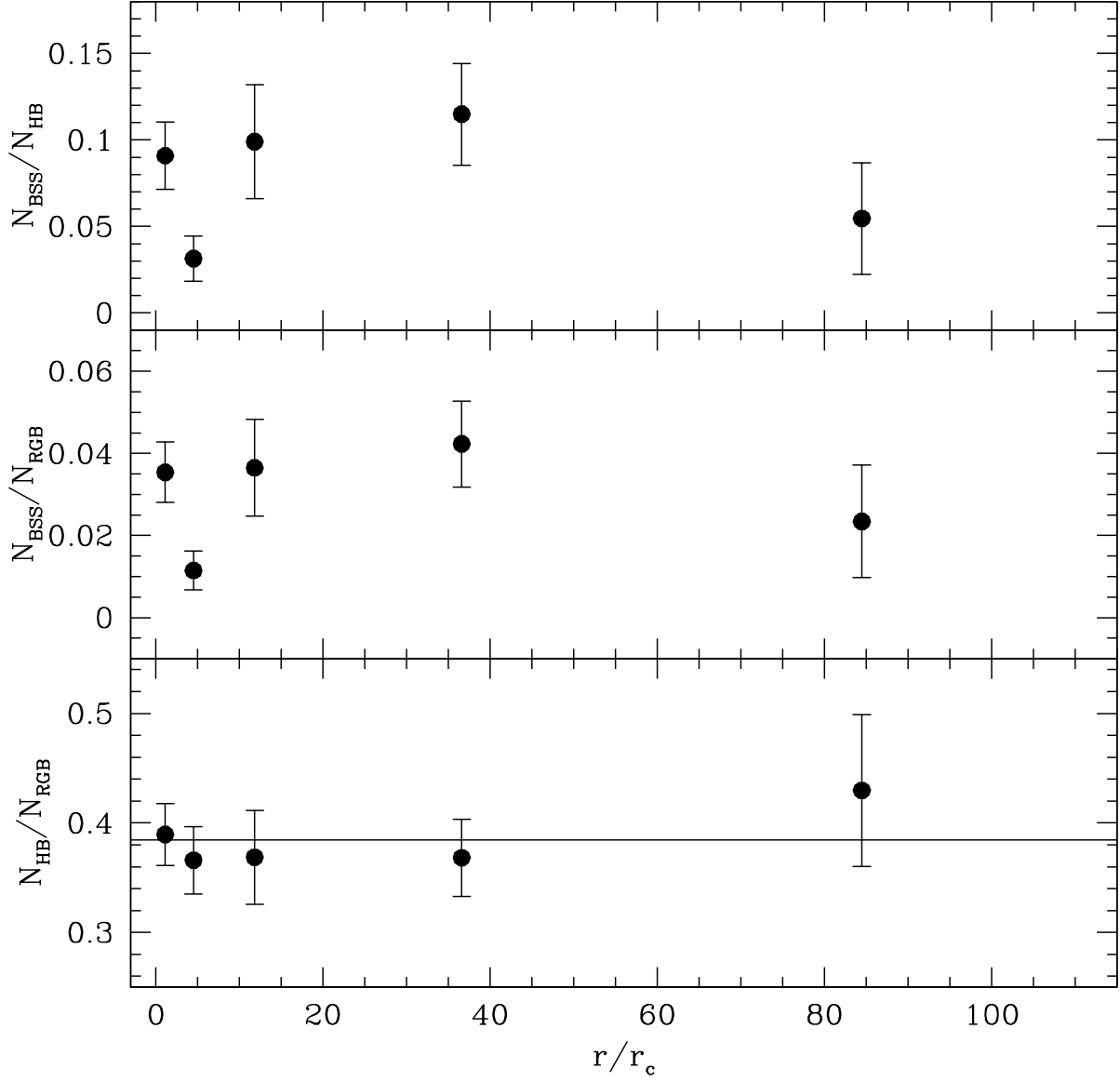}
\caption{Radial distribution of the population ratios $N_{\rm
    BSS}/N_{\rm HB}$, $N_{\rm BSS}/N_{\rm RGB}$, $N_{\rm HB}/N_{\rm
    RGB}$ (top, middle, and bottom panels, respectively) as a function
  of the radial distance from the cluster center, normalized to the
  core radius.}
\label{spec_freq}
\end{figure}

For a more quantitative analysis, we computed the population ratios
$N_{\rm BSS}/N_{\rm HB}$, $N_{\rm BSS}/N_{\rm RGB}$ and $N_{\rm
  HB}/N_{\rm RGB}$ in five concentric annuli centered in $C_{\rm grav}$. 
We adopted Poissonian errors for the populations and their propagation 
for the population ratios.
The star counts for each annulus are listed in Table 1.
Note that, due to the shape of the WFPC2 FOV, the annuli $10''<r<30''$
and $30''<r<75''$ are not entirely covered by our data (see Figure
\ref{map_hst}). The radial distributions of the population ratios are
shown in Figure \ref{spec_freq}.  The distribution of both $N_{\rm
  BSS}/N_{\rm HB}$ and $N_{\rm BSS}/N_{\rm RGB}$ is clearly bimodal,
with a peak in the center, a minimum located at $r_{min}\simeq
20\arcsec = 5 r_c$ and a rising branch in the outer regions. In
contrast, the $N_{\rm HB}/N_{\rm RGB}$ ratio (bottom panel) has a flat
behavior across the entire extension of the cluster, as expected for
normal populations following the distribution of the light.

We computed also the double normalized ratios for the three
populations, defined as (Ferraro et al. 1993, 1997):
\begin{equation}
R_{\rm pop}=\frac{N_{\rm pop}^{\rm ann}/N_{\rm pop}^{\rm tot}}{L^{\rm
    ann}_{\rm samp}/L^{\rm tot}_{\rm samp}}.
\end{equation}
We adopted Poissonian errors for the populations and the luminosity and their propagation 
for the double normalized ratios.

\begin{figure}
\includegraphics[scale=0.83]{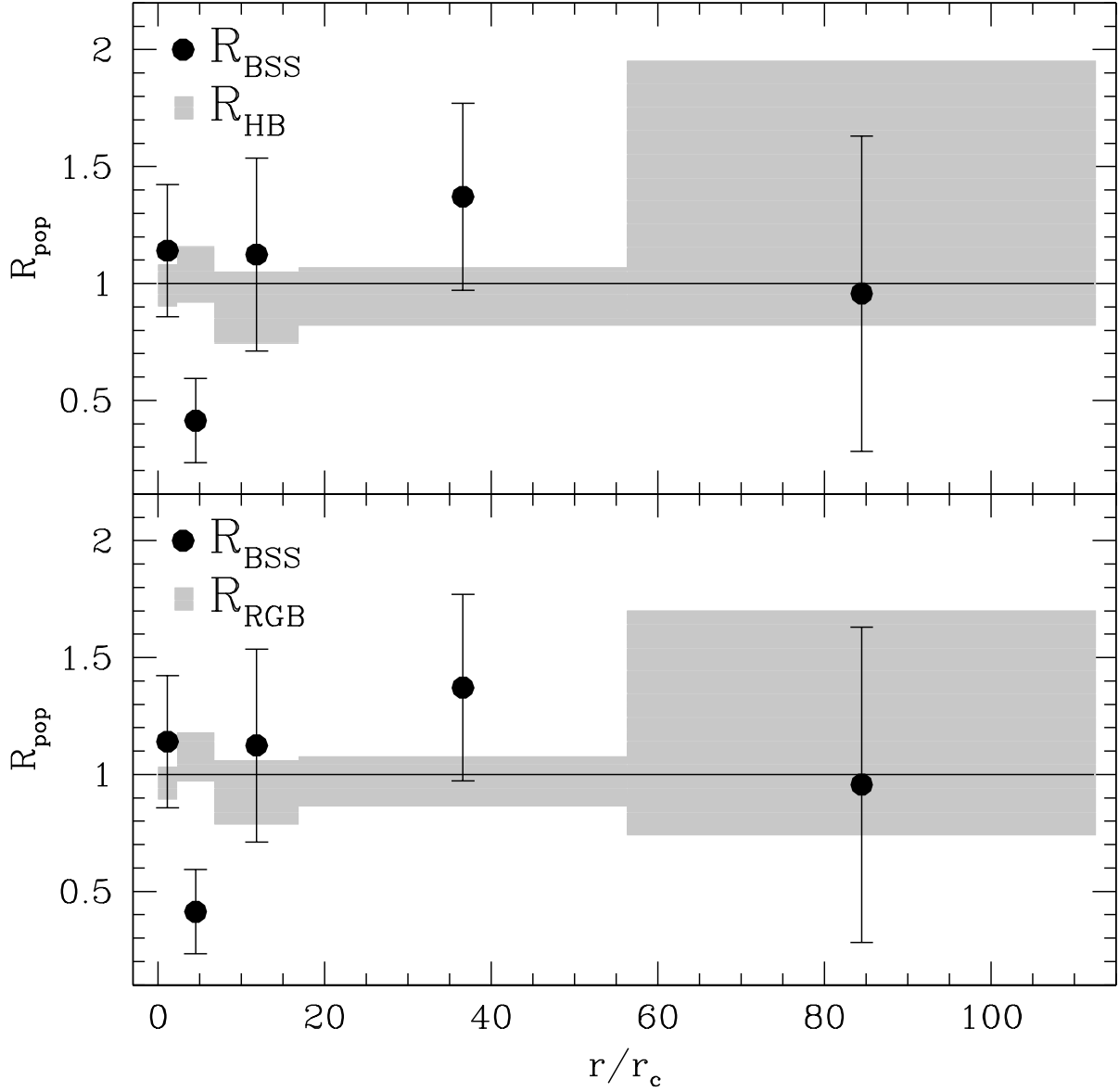}
\caption{Radial distribution of the double normalized ratios of BSSs
  (black dots) and the references stars (grey rectangles: HB stars in
  the top panel, RGB stars in the bottom one).}
\label{rpop}
\end{figure}

The luminosity in each annulus has been obtained by integrating the
best-fit Wilson profile and appropriately taking into account the
effective area covered by the observations. The computed luminosity
ratios are reported in Table 1. Figure \ref{rpop} shows the
results. As expected by the stellar evolutionary theory (Renzini \&
Fusi Pecci 1988), the HB and RGB stars follow the distribution of the
light: the radial distribution is constant (grey rectangles) at a
value close to unity. Again, the BSS distribution is found to be
bimodal.

\section{DISCUSSION}
Following the scenario proposed by F12, the shape of the BSS radial
distributions can be interpreted in terms of the dynamical age of
stellar systems. In particular, the bimodal BSS radial distribution
and the location of the minimum ($r_{min}\simeq 5 r_c$) found for NGC
5824 indicates that this cluster is a member of the intermediate
dynamical-age systems ({\it Family II}, in the F12 classification).
The BSS distribution is very similar to that found in M55 (see Figure
2 of F12), where the minimum is located at $r_{min}\sim 4 r_c$.  The
rising trend toward the center is very steep in both these clusters and
the minimum is well defined.  This means that dynamical friction has
already been effective in segregating BSSs toward the cluster center,
but it has poorly affected the outskirts.

\begin{figure}
\includegraphics[scale=0.83]{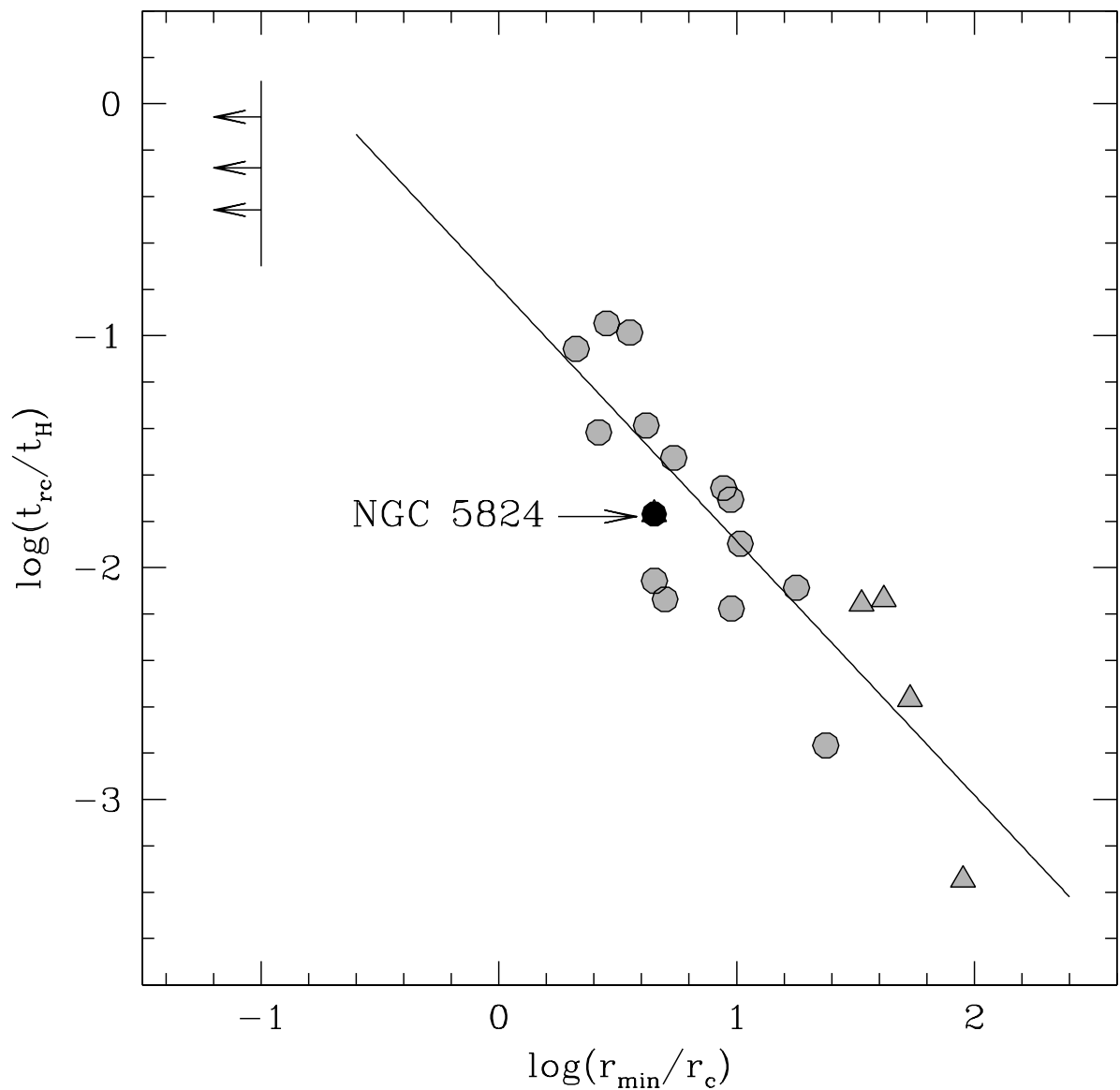}
\caption{Core relaxation time ($t_{rc}$) normalized to the age of the
  Universe ($t_{\rm H} = 13.7$ Gyr) as a function of $r_{min}/r_c$
  (from Figure 4 of F12).  The dynamically young systems (\emph{Family
    I}) are plotted as lower limit arrows at $r_{min}=0.1$. The grey
  triangles mark the dynamically old clusters (\emph{Family III}),
  while the grey circles mark the intermediate dynamical-age clusters
  (\emph{Family II}). The black circle corresponds to the position of
  NGC 5824 in this plane.}
\label{clock}
\end{figure}

As shown in F12, there is a strong relation between the core
relaxation time ($t_{rc}$, one of the classical theoretical indicator
of the cluster dynamical age) and the position of the minimum of the
BSS radial distribution measured in units of the cluster core radius
($r_{min}/r_c$). This allowed F12 to define the {\it empirical
  dynamical clock}: the position of $r_{min}/r_c$ can be used as a
sort of clock time-hand to measure the dynamical age of clusters. In
this framework, a flat BSS radial distribution, where $r_{min}$ cannot
be defined, indicates clusters with a relaxation time of the order of
the age of the Universe ({\it Family I}), while a monotonic BSS radial
distribution with only a central peak indicates dynamically old
clusters, where the action of dynamical friction has operated out to
the most remote regions of the cluster, segregating the entire BSS
population in the center ({\it Family III}).  Following F12, we have
computed $t_{rc}$ for NGC 5824 by using equation (10) of Djorgovsky
(1993), adopting the cluster structural parameters obtained in Section
3. We adopted the reddening and distance modulus quoted by Ferraro et
al. (1999), the central luminosity density listed by Harris (1996) and
the total cluster mass estimated by McLaughlin \& van der Marel
(2005).  In Figure \ref{clock} we show the position of NGC 5824 (black
solid circle) in the ``dynamical clock plane'' ($t_{\rm rc}/t_{\rm H}$
as a function of $r_{min}/r_c$; see Figure 4 in F12), where $t_{\rm
  H}= 13.7$ Gyr is the age of the Universe.  Clearly, this cluster
nicely follows the same relation defined by the sample analyzed in F12
(see also Dalessandro et al. 2013 and Beccari et al. 2013), further
confirming that the shape of the observed BSS distribution is a good
measure of GC dynamical ages. This figure also confirms that NGC 5824
is a mid dynamical-aged cluster of \emph{Family II}, where the action
of dynamical friction has already started to segregate BSSs (and
binary systems of similar total mass) toward the cluster center. In
this scenario, the most remote BSSs are thought to be still evolving
in isolation in the outer cluster regions.

\section{SUMMARY}
In this paper we have used a combination of HST UV and optical images
to sample the cluster center, and wide-field ground-based optical
observations covering the entire cluster extension to derive the main
structural parameters of the globular cluster NGC 5824 and to study
its BSS population.

From the high-resolution data we derived the cluster center of gravity 
lying at $\alpha (J2000.0) = 15^{\rm h} 3^{\rm m} 58.64^{\rm s}$,
$\delta (J2000.0) = -33^{\circ} 4' 5.905''$. We determined the radial
density profile from star counts, finding that it is best fitted by a
Wilson model with core radius $r_c\simeq4.4''$ and concentration
$c\simeq2.74$.

For the first time, the age of this cluster has been determined. Using
isochrone fitting we found $t=13\pm0.5$ Gyr, in excellent agreement
with what expected for a outer-halo Galactic GC.

A total of 60 BSSs (40 in the \textit{inner sample} and 20 in the
\textit{outer sample}) has been identified.  The comparison between
the radial distribution of BSSs and normal cluster stars (HB and RGB),
as well as of the double normalized ratios shows that the BSS
distribution is bimodal: peaked in the center, with a clear-cut dip at
intermediate-small radii ($r_{min}\simeq20''\simeq 5 r_c$), and with
an upturn in the external regions.  This suggests that NGC 5824 is an
intermediate dynamical-age cluster.

\section*{ACKNOWLEDGMENTS}
This research is part of a Project COSMIC-LAB founded by The European
Research Council (under contract ERC-2010-AdG-267675).  Research at
the University of Virginia was supported in part by NASA grant
GO-11975 from the Space Telescope Science Institute to R. T. Rood and
R. W. O'Connell.

\begin{table}
\begin{tabular}{|p{0.95cm}||p{0.95cm}||p{0.95cm}||p{0.95cm}||p{0.95cm}||p{0.95cm}||p{0.95cm}||p{0.95cm}||p{2.cm}|*{9}{c|}|}
\hline
$r''_i$&$r''_e$&$N_{\rm BSS}$&$N_{\rm BSS}^{\rm field}$&$N_{\rm HB}$&$N_{\rm HB}^{\rm field}$&$N_{\rm RGB}$&$N_{\rm RGB}^{\rm field}$&$L^{\rm ann}_{\rm samp}/L^{\rm tot}_{\rm samp}$\\
\hline
  0 &  10 & 24 & 0 & 264 &  0 & 678 &   0 & 0.35\\
 10 &  30 &  6 & 0 & 191 &  0 & 523 &   1 & 0.24\\
 30 &  75 & 10 & 0 & 102 &  1 & 277 &   3 & 0.15\\
 75 & 250 & 18 & 1 & 162 & 14 & 467 &  65 & 0.21\\
250 & 500 &  5 & 2 & 100 & 45 & 341 & 213 & 0.05\\
\hline
\end{tabular}
\caption{Internal and external radii of the five adopted annuli,
  number of observed BSSs, HB and RGB stars, estimated number of
  contaminating field stars for each populations, and fraction of
  sampled light.}
\end{table}

\label{lastpage}

\begin{thebibliography}{}

\bibitem[Bailyn(1992)]{ba92} Bailyn, C. D.\ 1992, ApJ, 392, 
519 

\bibitem[Bailyn (1995)]{ba95} Bailyn, C. D. 1995, ARA\&A, 33, 133

\bibitem[Beccari et al.(2011)]{2011ApJ...737..3B} Beccari, G.,
  Sollima, A., Ferraro, F. R., Lanzoni, B., Bellazzini, M., De Marchi,
  G., Valls-Gabaud, D., Rood, R. T. 2011, ApJ Letters, 737, 3

\bibitem[Beccari et al.(2013)]{bec13} Beccari, G., 
Dalessandro, E., Lanzoni, B., et al.\ 2013, ApJ, 776, 60 

\bibitem[Bellazzini et al.(1995)]{bella95} Bellazzini, M., Pasquali,
  A., Federici, L., Ferraro, F.~R., \& Pecci, F.~F.\ 1995, ApJ, 439,
  687

\bibitem[Bressan et al. (2012)]{bre12} Bressan, A., Marigo, P.,
  Girardi, L., Salasnich, B., Dal Cero, C., Rubele, S., Nanni,
  A. 2012, MNRARS, 427, 127

\bibitem[Brown et al.(2008)]{2008ApJ...682..319B} Brown, T.~M., Smith, E., 
Ferguson, H.~C., et al.\ 2008, ApJ, 682, 319

\bibitem[Contreras Ramos et al. (2012)]{cont12} Contreras Ramos, R.,
  Ferraro, F. R., Dalessandro, E., Lanzoni, B., Rood, R. T. 2012, ApJ,
  748, 91

\bibitem[Dalessandro et al.(2008b)]{2008ApJ...677.1069D} Dalessandro, E., 
Lanzoni, B., Ferraro, F.~R., et al.\ 2008b, ApJ, 677, 1069 

\bibitem[Dalessandro et al.(2008a)]{dal08} Dalessandro, E., 
Lanzoni, B., Ferraro, F.~R., Vespe, F., Bellazzini, M., 
\& Rood, R.~T.\ 2008a, ApJ, 681, 311

\bibitem[Dalessandro et al.(2011)]{2011MNRAS.410..694D} Dalessandro, E., 
Salaris, M., Ferraro, F.~R., et al.\ 2011, MNRAS, 410, 694  

\bibitem[Dalessandro et al.(2013)]{dal13} Dalessandro, E., 
Ferraro, F.~R., Lanzoni, B., et al.\ 2013, ApJ, 770, 45 

\bibitem[Dalessandro et al.(2013)]{dal13} Dalessandro, E., 
Ferraro, F.~R., Massari, D. et al.\ 2013, arXiv1310.2389 

\bibitem[D'Cruz et al.(2000)]{2000ApJ...530..352D} D'Cruz, N.~L., 
O'Connell, R.~W., Rood, R.~T., et al.\ 2000, ApJ, 530, 352

\bibitem[De Marco et al. (2004)]{dem04} De Marco, O., Lanz, T.,
  Ouellette, J., A. Zurek, D., Shara, M. M. 2004, ApJ, 606L, 151

\bibitem[Dotter et al. (2010)]{dot10} Dotter, A., Sarajedini, A., Anderson, J., et al.
2010, ApJ, 708, 698

\bibitem[Dotter et al. (2011)]{dot11} Dotter, A., Sarajedini, A., Anderson, J. 2011, ApJ, 738, 74

\bibitem[Dolphin (2009)]{dolph09}Dolphin, A.~E.\ 2009, PASP, 121, 655

\bibitem[Ferraro et al.(1991)]{fe91} Ferraro, F.~R., 
Clementini, G., Fusi Pecci, F., \& Buonanno, R.\ 1991, MNRAS, 252, 357 

\bibitem[Ferraro et al.(1992)]{fe92} Ferraro, F.~R., 
Fusi Pecci, F., \& Buonanno, R.\ 1992, MNRAS, 256, 376 

\bibitem[Ferraro et al.(1993)]{fe93} Ferraro, F.~R., Fusi Pecci, F.,
  Cacciari, C., Corsi, C. E., Buonanno, R., Fahlman, G. G., Richer,
  H. B. \ 1993, AJ, 106, 2324

\bibitem[Ferraro et al.(1995)]{fe95} Ferraro, F.~R., Fusi Pecci, F., Bellazzini, M. \ 1995,
A\&A, 294, 80 

\bibitem[Ferraro et al.(1997)]{fe97} Ferraro, F.~R., Paltrinieri, B., Fusi
Pecci, F., Cacciari, C., Dorman, B., Rood, R.~T., Buonanno, R., Corsi, C.~E.,
Burgarella, D., \& Laget, M. 1997, A\&A, 324, 915

\bibitem[Ferraro et al.(1998)]{1998ApJ...500..311F} Ferraro, F.~R.,
  Paltrinieri, B., Pecci, F.~F., Rood, R.~T., \& Dorman, B.\ 1998,
  ApJ, 500, 311

\bibitem[Ferraro et al.(1999)]{1999AJ....118.1738F} Ferraro, F.~R., 
Messineo, M., Fusi Pecci, F., et al.\ 1999, AJ, 118, 1738 

\bibitem[Ferraro et al.(2001)]{fe01} Ferraro, F.~R., D'Amico, N.,
  Possenti, A., Mignani, R.~P., \& Paltrinieri, B.\ 2001, ApJ, 561,
  337

\bibitem[Ferraro et al.(2003)]{2003ApJ...595..179F} Ferraro, F.~R.,
  Possenti, A., Sabbi, E., et al.\ 2003, ApJ, 595, 179

\bibitem[Ferraro et al.(2006a)]{fe06a} Ferraro, F.~R., Sabbi, E.,
  Gratton, R. et al.\ 2006a, ApJ, 647, L53

\bibitem[Ferraro et al.(2006b)]{fe06b} Ferraro, F.~R., Sollima, A., Rood,
R.~T., Origlia, L., Pancino, E., \& Bellazzini, M. \ 2006b, ApJ, 638, 433

\bibitem[Ferraro et al.(2009)]{fe09} Ferraro, F.~R., Beccari, G., Dalessandro, E., et al.
\ 2009, Nature, 462, 1028

\bibitem[Ferraro et al. (2012)]{fe12} Ferraro, F. R., Lanzoni, B.,
  Dalessandro, et al.
2012, Nature, 492, 393

\bibitem[Georgiev et al. (2009)]{geo09} Georgiev, I. Y., Hilker, M.,
  Puzia, T. H., Goudfrooij, P., Baumagrdt 2009, MNRAS, 396, 1075

\bibitem[Gilliland et al. (1998)]{gil98} Gilliland, R. L., Bono, G.,
  Edmonds, P. D., Caputo, F., Cassisi, S., Petro, L. D., Saha, A.,
  Shara, M. M.  1998, ApJ, 507, 818

\bibitem[Grillmair et al. (1995)]{gri95} Grillmair, C. J., Freeman,
  K. C., Irwin, M., Quinn, P. J.1995, AJ, 109, 2553

\bibitem[Harris (1996)]{har96} Harris, W.E. 1996, AJ, 112, 1487 (2010
  version)

\bibitem[Hills\&Day (1976)]{hill76} Hills, J. G., Day, C. A. 1976, ApL, 17, 87

\bibitem[Holtzman et al. (1995)]{holtz95} Holtzman, J.~A., Burrows, C.~J.,
Casertano, S., Hester, J.~J., Trauger, J.~T., Watson, A.~M., \& Worthey, G. \
1995, PASP, 107, 1065

\bibitem[Hut et al.(1992)]{hut92} Hut, P., McMillan, S., Goodman, J.,
  et al.\ 1992, PASP, 104, 981

\bibitem[King (1966)]{ki66} King, I. R. 1966, AJ, 71, 64

\bibitem[Knigge et al. (2009)]{kni09} Knigge, C., Leigh, N., Sills, A. 2009, Nature, 457, 288

\bibitem[Lanzoni et al. (2007a)]{lan07_1904} Lanzoni, B., Sanna, N.,
  Ferraro, et al. \ 2007a ApJ, 663, 1040

\bibitem[Lanzoni et al.(2007b)]{2007ApJ...663..267L} Lanzoni, B.,
  Dalessandro, E., Ferraro, F.~R., et al.\ 2007b, ApJ, 663, 267

\bibitem[Li et al. 2013]{Li13} Li, C., de Grijs, R. Deng, L., Liu, X. K. 2013, ApJ, 770, L7

\bibitem[Lovisi et al.(2013)]{2013ApJ...772..148L} Lovisi, L.,
  Mucciarelli, A., Lanzoni, B., et al.\ 2013, ApJ, 772, 148

\bibitem[L{\"u}tzgendorf et al.(2013)]{2013A&A...552A..49L}
  L{\"u}tzgendorf, N., Kissler-Patig, M., Gebhardt, K., et al.\ 2013,
  A\&A, 552, A49

\bibitem[Mapelli et al. (2004)]{map04} Mapelli, M., Sigurdsson, S., Colpi, M.,
Ferraro, F.~R., Possenti, A., Rood, R.~T., Sills, A., Beccari, G. 2004, ApJ
Letters, 605, 29 

\bibitem[McCrea (1964)]{mtbss1} McCrea, W.~H.\ 1964, MNRAS, 128, 147

\bibitem[McLaughlin \& van der Marel(2005)]{mcL05} McLaughlin, D.~E., \& van
der Marel, R.~P. \ 2005, ApJs, 161, 304

\bibitem[Meylan \& Heggie(1997)]{me97} Meylan, G., \& 
Heggie, D.~C.\ 1997, Astron. Astrophys. Rev., 8, 1 

\bibitem[Miocchi et al.(2013)]{2013ApJ...774..151M} Miocchi, P.,
  Lanzoni, B., Ferraro, F.~R., et al.\ 2013, \apj, 774, 151

\bibitem[Montegriffo et al. (1995)]{mont95} Montegriffo, P., Ferraro, F.~R.,
Fusi Pecci, F., \& Origlia, L.\	1995, MNRAS, 276, 739

\bibitem[Newberg et al. (2009)]{new09} Newberg, H. J., Yanny, B.,
  Willett, B. A. 2009, ApJ, 700, 61

\bibitem[O'Connell et al.(1997)]{1997AJ....114.1982O} O'Connell, R.~W., 
Dorman, B., Shah, R.~Y., et al.\ 1997, AJ, 114, 1982 

\bibitem[Paresce et al. (1991)]{pare91} Paresce, F., Meylan, G.,
  Shara, M., Baxter, D., \& Greenfield, P. 1991, Nature, 352, 297

\bibitem[Paresce et al.(1992)]{pare92} Paresce, F., de Marchi, 
G., \& Ferraro, F.~R.\ 1992, Nature, 360, 46 

\bibitem[Pooley 
\& Hut(2006)]{pool06} Pooley, D., \& Hut, P.\ 2006, ApJL, 646, L143 

\bibitem[Ransom et al.(2005)]{ran05} Ransom, S.~M., Hessels, 
J.~W.~T., Stairs, I.~H., et al.\ 2005, Science, 307, 892 

\bibitem[Renzinj et al. (1988)]{re88} Renzini, A., Fusi Pecci, F. 1988, ARA\&A, 26, 199

\bibitem[Samus et al (2009)]{sam09} Samus, N. N., Kazarovets, E. V.,
  Pastukhova, E. N., Tsvetkova, T. M., Durlevich, O. V.  2009, PASP,
  121, 1378

\bibitem[Sanna et al. (2012)]{san12} Sanna, N., Dalessandro, E.,
  Lanzoni, B., Ferraro, F. R., Beccari, G., Rood, R. T. 2012, MNRAS,
  422, 1171

\bibitem[Sarna \& De Greve (1996)]{sar96} Sarna, M. J., De Greve, J. -P. 1996, QJRAS, 37, 11

\bibitem[Saviane et al. (2012)]{sav12} Saviane, I., da Costa, G. S.,
  Held, E. V., Sommariva, V., Gullieuszik, M., Barbury, B.,
  Ortolani. S. 2012, A\&A, 540, 27

\bibitem[Schiavon et al.(2012)]{2012AJ....143..121S} Schiavon, R.~P., 
Dalessandro, E., Sohn, S.~T., et al.\ 2012, AJ, 143, 121 

\bibitem[Shara et al. (1997)]{sha97} Shara, M.~M., Saffer, R.~A., \& Livio,
M.\ 1997, ApJ, 489, L59

\bibitem[Stetson(1987)]{dao} Stetson, P.~B.\ 1987, PASP, 
99, 191 

\bibitem[Stetson(1994)]{dao2} Stetson, P.~B.\ 1994, PASP, 
106, 250 

\bibitem[Stetson(2000)]{dao3} Stetson, P.~B.\ 2000, PASP, 
112, 925

\bibitem[Valenti et al.(2004)]{valenti04} Valenti, E.,
  Ferraro, F.~R., \& Origlia, L.\ 2004, MNRAS, 351, 1204

\bibitem[Wilson (1975)]{wil75} Wilson, C. P. 1975, AJ, 80, 175

\bibitem[Zinn\&Searle (1976)]{zin76} Zinn, R., Searle, L. 1976, ApJ, 209, 734

\end{thebibliography}
\end{document}